\documentclass[12pt]{article}

\pdfoutput=1

\usepackage{amsmath,amssymb,graphicx} 
\usepackage{epsf}
\usepackage{pstricks}
\usepackage{cite}

\newcommand{\beq}{\begin{eqnarray}}
\newcommand{\eeq}{\end{eqnarray}}
\newcommand{\be}{\begin{equation}}
\newcommand{\ee}{\end{equation}}

\newcommand{\solar}{\odot}
\newcommand{\alphabar}{\bar{\alpha}}
\newcommand{\alphaem}{\alpha_{em}}
\newcommand{\Eres}{{E_\gamma^{\rm res}}}
\newcommand{\rb}{r_B}
\newcommand{\ra}{\rightarrow}

\newcommand{ \centeron }[2]{{\setbox0=\hbox{#1}\setbox1=\hbox{#2}\ifdim
                             \wd1>\wd0\kern.5\wd1\kern-.5\wd0\fi \copy0
                             \kern-.5\wd0\kern-.5\wd1\copy1\ifdim\wd0>\wd1
                             \kern.5\wd0\kern-.5\wd1\fi}}
\newcommand{ \ltap }{\>\centeron{\raise.35ex\hbox{$<$}}
                     {\lower.65ex\hbox{$\sim$}}\>}
\newcommand{ \gtap }{\>\centeron{\raise.35ex\hbox{$>$}}
                     {\lower.65ex\hbox{$\sim$}}\>}
\newcommand{ \gsim }{\mathrel{\gtap}}
\newcommand{ \lsim }{\mathrel{\ltap}}

\begin{document}
\begin{titlepage}

\vspace*{0.5cm}

\centerline{\LARGE\bf Quirky Composite Dark Matter}

\vspace*{0.5cm}

\begin{center}
Graham D. Kribs$^a$, 
Tuhin S. Roy$^a$, 
John Terning$^b$, and 
Kathryn M. Zurek$^{c,d}$
\end{center}

\vspace*{0.5cm}

\noindent
\hspace*{0.5cm}
$^a${\it Department of Physics, University of Oregon, Eugene, OR, 97403} \\
\hspace*{0.5cm}
$^b${\it Department of Physics, University of California, Davis, CA 95616} \\
\hspace*{0.5cm}
$^c$\mbox{\it Particle Astrophysics Center, 
              Fermi National Accelerator Laboratory, Batavia, IL 60510} \\
\hspace*{0.5cm}
$^d${\it Department of Physics, University of Michigan, Ann Arbor, MI 48109} \\

\vspace*{0.4cm}

\begin{abstract}
\noindent

We propose a new dark matter candidate, ``quirky dark matter,'' 
that is a scalar baryonic bound state of a new non-Abelian 
force that becomes strong below the electroweak scale.
The bound state is made of chiral quirks: new fermions that transform 
under both the new strong force as well as in a chiral representation 
of the electroweak group, acquiring mass from the Higgs mechanism.
Electric charge neutrality of the lightest baryon requires 
approximately degenerate quirk
masses which also causes the charge radius of the bound state 
to be negligible.  
The abundance is determined by an asymmetry that is linked to
the baryon and lepton numbers of the universe through 
electroweak sphalerons. 
Dark matter elastic scattering with nuclei proceeds through 
Higgs exchange as well as an electromagnetic polarizability operator
which is just now being tested in direct detection experiments.
A novel method to search for quirky 
dark matter is to look for a gamma-ray ``dark line'' spectroscopic 
feature in galaxy clusters that result from the quirky Lyman-alpha or
quirky hyperfine transitions.  Colliders are expected to dominantly 
produce quirky mesons, not quirky baryons, consequently large missing 
energy is not the primary collider signal of the physics associated 
with quirky dark matter.

\end{abstract}

\end{titlepage}

\newpage
\setcounter{page}{2}

\section{Introduction}
\label{sec:intro}
\setcounter{equation}{0}

Dark matter is a big mystery.  Weakly interacting massive particles
(WIMPs) provide one interesting class of particles to serve as dark matter.
There are, nevertheless, two main puzzles with typical WIMP candidates:

\begin{itemize}
\item[(1)] Abundance is determined by thermal freezeout.
While a thermal freezeout origin can yield the observed 
cosmological abundance when the annihilation cross section 
is tuned to roughly $1$ pb, this mechanism is entirely unrelated 
to the origin of matter, which arises from an asymmetry.  
The observational relation between the dark matter density and
the baryonic density, $\rho_{D} \simeq 5 \rho_{B}$, 
is a coincidence.
\item[(2)] Elementary WIMPs with electroweak interaction strength couplings 
to standard model (SM) fermions generically have very strong constraints 
from direct detection bounds.  To be ``safe'', WIMP interactions with the 
SM must be sub-weak strength, and typically, their mass determined
by a mechanism unrelated to electroweak symmetry breaking.
\end{itemize}

We propose a new model of dark matter that tackles both problems.
The first problem can be addressed if the dark matter abundance 
is linked to the baryon abundance.  This has been considered before, 
for example, in the context of technibaryon dark matter
\cite{Chivukula:1989qb,Barr:1990ca,Chivukula:1992pn,Bagnasco:1993st,Gudnason:2006yj,Foadi:2008qv}.  
In these models, electroweak sphalerons re-process baryon and lepton 
asymmetries into technibaryon asymmetry.  The constituents are 
electroweak charged, while the technibaryon composite dark matter 
is neutral.  The sphalerons generate a relation between the 
number densities of leptons, baryons and technibaryons,
\begin{equation}
n_{\ell} - n_{\bar{\ell}} \sim n_b - n_{\bar{b}} \sim n_{tb} - n_{\bar{tb}},
\label{proportions}
\end{equation}
where $n_{\ell} - n_{\bar{\ell}}$, $n_b - n_{\bar{b}}$, and 
$n_{tb} - n_{\bar{tb}}$ represent the lepton, baryon and 
technibaryon asymmetries, and the exact proportions are ${\cal O}(1)$ 
depending on the electroweak charges of the technibaryon constituents.  
Cosmologically $\rho_{D}/\rho_{B} \approx 5$, so that the relation 
Eq.~(\ref{proportions}) implies the dark matter mass 
$M \sim 5$~GeV\@.  If this were the end of the story, 
technibaryon dark matter (or any other weak scale model 
which connects the dark matter to the baryon asymmetry, 
e.g.\ \cite{Kaplan:1991ah,Banks:2005hc}) would be ruled out.  
However, if the dark matter constituents are just becoming 
non-relativistic as the sphalerons are decoupling, there is 
an exponential Boltzmann suppression in the technibaryon asymmetry 
relative to the lepton and baryon asymmetries, implying the 
dark matter can naturally have an electroweak scale mass.  
The other possible solution to the dark matter baryon coincidence 
places the GeV scale dark matter in a hidden sector weakly coupled 
to the SM sector \cite{Fujii:2002aj,Hooper:2004dc,Kitano:2004sv,Cosme:2005sb,Farrar:2005zd,Suematsu:2005zc,Kitano:2008tk,Seto,Kaplan:2009ag}.

The second problem can, in fact, be ingeniously solved by 
compositeness.  In technicolor theories, technifermions
interact with the SM through renormalizable interactions, 
while a composite technibaryon dark matter candidate is charge- 
and electroweak-neutral.  This eliminates renormalizable interactions
with the SM below the electroweak scale, leaving only higher 
dimensional operators \cite{Chivukula:1992pn,Bagnasco:1993st}.
In the early 1990s this was thought to be an unfortunate result -- dark 
matter could not be observed in the-then foreseeable future.
Given the direct detection bounds today 
(e.g.\ \cite{Angle:2007uj,Ahmed:2008eu}), 
this becomes a ``feature'' of our composite dark matter model that 
we exploit to naturally suppress the direct detection cross sections.

The residual electroweak-mediated direct detection cross section
of composite dark matter arises from ``form factor'' suppression,
somewhat analogous to the suppression of the photon coupling to 
neutrons at energies much smaller than $\Lambda_{\rm QCD}$.  
For example, the leading order operators that couple scalar 
technibaryon dark matter to the SM arise at dimension-6 
(charge radius) \cite{Bagnasco:1993st} and 
dimension-7 (chromomagnetic polarizability)
\cite{Chivukula:1992pn}, suppressed by two or three powers 
of $\Lambda_{\rm TC}$.

What was not fully appreciated in the 1990s is that both of these 
operators can be eliminated.  The charge radius vanishes in a limit 
in which the current masses of the constituents are equal.  
The chromomagnetic polarizability vanishes when the constituents
do not carry ordinary QCD color.  While this suggests rethinking
technicolor dark matter (e.g.\ \cite{Gudnason:2006yj,Foadi:2008qv}), 
the model building difficulties of realizing a fully successful 
technicolor model incorporating flavor as well as avoiding electroweak 
precision constraints remains daunting.

In this paper, we take a different approach, in the spirit of the 
Hidden Valley \cite{Strassler:2006im} and
Kang and Luty's quirks \cite{Kang:2008ea}.  We retain the new 
strong dynamics, but discard their role in electroweak symmetry breaking.  
The new strong dynamics gets strong at a scale below the electroweak
scale.  We call the candidate that arises in this approach 
Quirky Dark Matter (QDM).
The simplest quirky dark matter model, as we will describe, 
contains exactly the same gauge and matter content as that of 
minimal SU(2) technicolor with two flavors.  Amusingly,
what was originally a problem of minimal technicolor --
namely the possibility that the vacuum aligned to an 
electroweak-preserving state \cite{Preskill:1980mz} -- is now a 
``feature'' here since we utilize the ordinary Higgs mechanism
to break electroweak symmetry.  Indeed, we do not want the 
strong dynamics to break electroweak symmetry (even a little bit)
lest we run into electroweak precision constraints.

With an ordinary Higgs present, quirks can obtain their mass through 
the Higgs mechanism just like quarks and leptons.  This has several
implications:  New contributions to the electroweak oblique 
corrections arise; we estimate them below.  Quirky dark matter
can interact with nuclei of direct detection experiments
through Higgs exchange; this leads to an ordinary elastic scattering
cross section that is right near the current bounds for a light
Higgs boson.  Finally, assuming the new strong force confines, 
new operators involving the Higgs are present that can allow 
the new glueballs to decay, independent of the quirk mass.

\section{Model}
\setcounter{equation}{0}

\subsection{Field Content}
\label{field-content-sec}

The model of quirky dark matter that we mainly wish to consider
consists of two flavors of quirks transforming under a new strongly 
interacting sector, $SU(2)_Q$, that hereafter we call ``quirkcolor.''
Variations of this, with different numbers of quirk flavors 
and quirkcolors are also possible; we will remark on the
possibility of more quirk flavors later in the paper.
We assume that the quirkcolor coupling constant gets strong 
\emph{below} the weak scale.  The particle content and charges under
the gauge and global symmetries are given in Table~\ref{table:model} 
in terms of two-component spinors.
\begin{table}
\begin{center}
\begin{tabular}{c|cccc}
        & $SU(2)_{Q}$ & $SU(2)_L$ & $U(1)_Y$ & $U(1)_{QB}$ \\ \hline
$\xi_Q = \begin{pmatrix} \xi_U, \xi_D\end{pmatrix}$     &  
        $\mathbf{2}$      &   $\mathbf{2}$     &   $0$  &  $+1/2$  \\
$\xi_{\bar{U}}$ &     $\mathbf{2}$      &    -      &   $-1/2$ & $-1/2$ \\
$\xi_{\bar{D}}$ &     $\mathbf{2}$      &    -      &   $+1/2$ & $-1/2$ 
\end{tabular}
\end{center}
\caption{Particle content and charges under the gauge and global symmetries.}
\label{table:model}
\end{table}
This assignment is chiral under the electroweak group, and thus 
requires Yukawa interactions with the Higgs,
\begin{eqnarray}
  \mathcal{L}_Y &=& \lambda_U \xi_Q H \xi_{\bar{U}} + 
             \lambda_D \xi_Q H^\dagger \xi_{\bar{D}}
\end{eqnarray}
to give current masses to the quirks, $m_q = \lambda_q v$,
for $q = U,D$.
We enforce a global $U(1)_{QB}$ ``quirky baryon number'' that forbids 
the mass terms $\xi_Q \xi_Q$ and $\xi_{\bar U} \xi_{\bar D}$
and ensures our quirky dark matter candidate is stable 
(on at least cosmological timescales). 

Since QDM contains additional matter that acquires mass exclusively 
through the Higgs mechanism, there are new contributions to 
the electroweak precision parameters.  
The quirks in our model are weakly-coupled at the scale of 
their mass, and so we can perturbatively calculate $\Delta S, \Delta T$
\cite{Peskin:1991sw}.  Generically, $\Delta S = 1/(3 \pi) \simeq 0.1$, 
while $\Delta T$ depends on the 
splitting within the quirky electroweak doublets.  
As we will show below, eliminating the charge radius operator 
requires negligible splitting 
between the current masses of the quirks.  As a consequence, 
the contribution to $T$ from this sector is negligible.  
The minimal model therefore has $\Delta S \simeq 0.1, \Delta T \simeq 0.0$, 
which is roughly at the 95\% CL contour when comparing against 
LEP electroweak working group fits \cite{Kribs:2007nz,Amsler:2008zzb}.
Suffice to say it is a straightforward (but unenlightening)
exercise to slightly extend the model to give a additional 
contributions to $T$ (and $S$) that result in a model fully consistent 
with electroweak precision data.

In the minimal model, with the only particles transforming under
$SU(2)_Q$ given by that in Table~\ref{table:model}, the
quirkcolor group confines.  The global $SU(4) \sim SO(6)$ symmetry 
is broken to $SO(5)$ and we have $15-10=5$ pseudo-Nambu-Goldstone bosons.
The large current quirk masses ensure these composites are massive,
forming ``quirkonia'' bound states with a spectrum similar to
heavy quarkonia \cite{Cheung:2008ke}.  We assume the quirkonia masses 
are sufficiently heavy to avoid LEP II bounds (i.e., quirks heavier 
than about 100 GeV).  Existing Tevatron bounds, as well as prospects 
for future detection at Tevatron and LHC, will be studied in a future
paper.

Confinement of $SU(2)_Q$ leads to quirkcolor glueballs.
These glueballs decay through higher dimensional operators
into SM matter.  Depending on the scales, however, their lifetime
may be very long \cite{Kang:2008ea,Juknevich:2009ji}, potentially
leading to cosmological problems depending on the quirk masses
and the confinement scale.  In QDM, there are additional operators 
due to interactions with the Higgs.  These interactions are written
with an estimate of their contribution to the glueball width 
in Sec.~\ref{sec:glueballdecay}.

Our quirkcolor group $SU(2)_Q$, however, does not necessarily 
need to confine, if additional (massless) flavors transforming only
under $SU(2)_Q$ are present.  This provides an interesting possibility 
in which quirkcolor flows to a conformal field theory \emph{without} 
confinement.  We emphasize that, for much of our discussion below, 
essentially none of our calculations depend on the scale (or existence) 
of confinement, so long as it is sufficiently smaller than the 
inverse Bohr radius of the bound states so that reliable non-relativistic 
calculations can be performed.  To this end, we need the 
quirkcolor coupling, 
evaluated at the scale of the inverse Bohr radius, to be perturbative.
The situation is quite analogous to heavy quarkonia.  Indeed, we employ
much of the formalism of non-relativistic effective theories 
developed for quarkonia and apply it directly to QDM\@.
The systematic derivation of the non-relativistic limit from the 
relativistic degrees of freedom, following the quarkonia literature,
is outlined in Appendix~\ref{NR-app}.

\begin{figure}
\setlength{\unitlength}{1mm}
\begin{center}
\begin{picture}(130,70)(0,0)
\put(20,10){\line(1,0){15}}
\put(20,30){\line(1,0){15}}
\put(45,30){\line(1,0){15}}
\put(80,12){\line(1,0){15}}
\put(80,32){\line(1,0){15}}
\put(105,32){\line(1,0){15}}
\multiput(40,10)(2,0){18}{\line(1,0){1}}
\multiput(40,12)(2,0){18}{\line(1,0){1}}
\put(5,10){\makebox(15,0)[l]{$n=1$}}
\put(5,30){\makebox(15,0)[l]{$n=2$}}
\put(5,50){\makebox(15,0)[l]{$n=3 \;\;\; \cdots$}}
\put(20,10){\makebox(15,5)[c]{$s$}}
\put(20,30){\makebox(15,5)[c]{$s$}}
\put(45,30){\makebox(15,5)[c]{$p$}}
\put(80,12){\makebox(15,5)[c]{$s$}}
\put(80,32){\makebox(15,5)[c]{$s$}}
\put(105,32){\makebox(15,5)[c]{$p$}}
\put(40,50){\makebox(0,0)[c]{$\vdots$}}
\put(100,50){\makebox(0,0)[c]{$\vdots$}}
\put(40,0){\makebox(0,70)[t]{$B^0_0$}}
\put(40,0){\makebox(0,65)[t]{(spin-singlet)}}
\put(100,0){\makebox(0,70)[t]{$B^0_1$}}
\put(100,0){\makebox(0,65)[t]{(spin-triplet)}}
\put(47,27){\vector(-1,-1){15}}
\put(45,20){\makebox(40,0)[l]{quirky Lyman-alpha}}
\put(57,12){\vector(0,-1){2}}
\put(62,5){\line(-1,1){5}}
\put(65,3){\makebox(40,0)[l]{quirky hyperfine}}

\end{picture}
\end{center}
\caption{Sketch of the quantum mechanical energy spectrum of 
our quirky dark matter composite with the ground state and 
several excited states shown.
Our notation $B^q_S$ corresponds to baryonic states with total 
electric charge $q$ and total spin $S$.  We have included 
$\mathcal{O}(\alphabar^2)$ (quirky Lyman-alpha) and 
$\mathcal{O}(\alphabar^4)$ (quirky hyperfine) splittings,
but do not show the $\mathcal{O}(\alphabar^5)$ (quirky Lamb shift)
splittings or other higher-order effects.  
The lightest electrically charged baryons $B^\pm_1$ (not shown),
have spin one, and are slightly heavier than $B^0_1$ due to 
subdominant electromagnetic corrections to the potential.}
\label{spectrum-fig}
\end{figure}
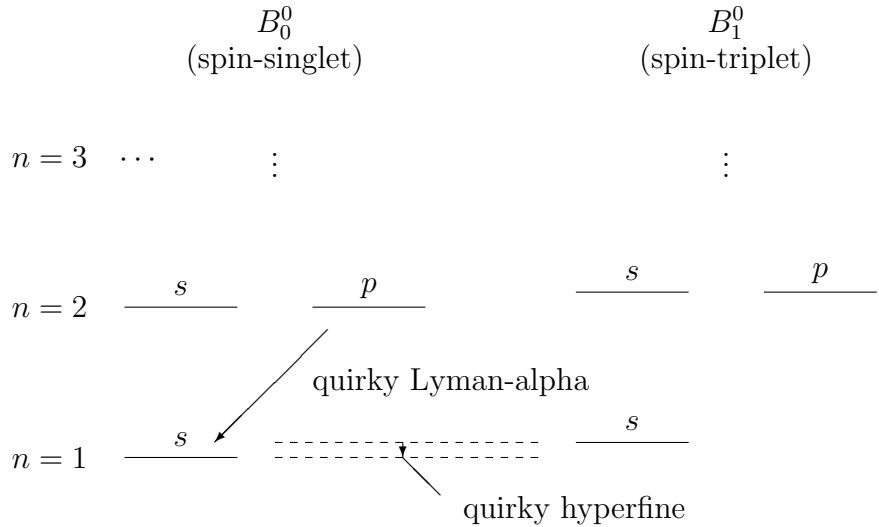

Our composites include ``mesons'' and ``baryons'' depending on whether 
or not they carry the nonzero $U(1)_{QB}$ quirky baryon number.  
To satisfy the Pauli exclusion principle, the wavefunction of 
baryons must be anti-symmetrized with respect to all quantum numbers 
leading to a detailed spectrum of allowed states.  Details outlining the
construction of the bound state spectrum from the relativistic spinors 
to the non-relativistic mesons and baryons are given in 
Appendix~\ref{NR-app}.  Here we simply quote the results and present 
the baryon spectrum that is relevant for dark matter and its interactions, 
given in Figure~\ref{spectrum-fig}.  

The dynamics and binding energies of the quirky baryons is our
primary interest in this paper.  For this, we need to construct 
the non-relativistic potential, $V_s(r)$.  Formally, this is possible 
in the limit $m_q \gg m_q v \gg  m_q v^2$, where the potential is dominated 
by single quirkcolor gauge field exchange between the massive 
quirks\footnote{If $SU(2)_Q$ confines at a scale $\Lambda_\text{Q}$, 
we actually need $m_q v \gg  m_q v^2 \gtrsim \Lambda_\text{Q}$.  
This is because
in the alternate limit $m_q v \gg \Lambda_\text{Q} \gg  m_q v^2$ 
one must integrate out the physical scale $\Lambda_\text{Q}$ 
before the potential can be properly defined.  This procedure 
leads to an additional non-perturbative part in the potential 
that contains non-local quirkcolor gauge field correlators.}.
Since our binding quirkcolor force is non-Abelian, the strength of 
the potential depends on the representation of the constituents.   
For our model, the quirks are in the fundamental representation,
which gives a potential
\begin{eqnarray}
  V_s(r) &=& - \, \frac{\alphabar(r)}{r}\; ,
\label{potn}
\end{eqnarray}
where\footnote{We drop the subscript $Q$ when writing $\alphabar$ 
to keep the notation as simple as possible.}
$\alphabar(r) \equiv C_2(\mathbf{2}) \, \alpha_Q(r) =\frac{3}{4} \alpha_Q(r)$,
and $\alpha_Q(r)$ is the strength of the quirkcolor force
evaluated at the scale $1/r$.
This potential strictly arises from the one quirkcolor gluon 
exchange in the singlet channel, 
$(\sum_a t^a_{ij} t^a_{kl}) \delta_{jl} = C_2(\mathbf{2}) \delta_{ij}$.
This potential is similar to the Coulombic potential used to
determine the bound states of the hydrogen atom.  
However, the non-Abelian nature of the quirkcolor binding force
implies $\alpha_Q(r)$ is scale-dependent, which to leading-log is
given by the $\beta$-function\cite{Fischler:1977yf,Billoire:1979ih}
\begin{equation}
  \alpha_Q(r) = \alpha_Q( \rb ) \left(  1 + \frac{\alpha_Q( \rb )}{3 \pi} 
      (11 - N_f) \ln(r/\rb) \right)\; .
\label{alpha-run}
\end{equation}
Here $\rb \equiv [\alphabar(\rb) \mu]^{-1}$ is the analogue of 
the Bohr radius in the hydrogen atom and gives the typical size 
of the bound states.  We have written the $r$-dependent correction
to the potential for a general number of quirkcolor flavors
for completeness.  The model on which we concentrate our attention
has $N_f = 2$, as defined before in Table~\ref{table:model}.

\subsection{Binding energies and splittings}
\label{binding-sec}

Our quirkcolor singlet bound states 
can be described by a Schr\"odinger-like equation with a quirkcolor
force potential given by Eqs.~(\ref{potn}),(\ref{alpha-run}).  
In the limit that the $\log r$ piece can be neglected, 
the potential becomes purely Coulombic -- the same as the 
hydrogen atom -- with the replacements $m_e \leftrightarrow \mu$ and 
$\alphaem \leftrightarrow \alphabar$.  
One can formally approach the Coulombic limit if enough additional 
flavors are present to lead to a nearly scale-invariant 
quirkcolor theory while $\alphabar$ remains perturbative.

We will be interested in the regime where $\alphabar$ is perturbative 
but not necessarily small, and with exactly two flavors of
quirks as given in Table~\ref{table:model}.  Hence, our non-relativistic 
potential has unavoidable $\log r$ dependence.  
In our calculations below, we express 
the effect of the $\log$ term as coefficients that multiply the 
exact solutions obtained in the case of a purely Coulombic potential.  
The coefficients have been computed by numerically solving the 
Schr\"odinger equation including the $\log$ term for a few specific 
choices of $\alphabar$.

The binding energies of the $n$-th excited state of the quirkcolor
singlets is given by
\begin{eqnarray}
E_n &=& -  k_n \frac{\alphabar^2 \mu }{2 \, n^2} \; ,
\label{Lyman-eq}
\end{eqnarray}
expressed in terms of the reduced mass of the quirks, 
$1/\mu \equiv 1/m_U + 1/m_D$.  The $n$-dependent constant 
$k_n$ captures the difference
between a pure Coulombic potential and our non-Abelian quirkcolor theory.
Using our numerical evaluation of the Schr\"odinger equation, 
we find the energy levels of the first two states are corrected by 
$k_1 \simeq (1.2,1.3,1.4)$ and $k_2 \simeq (1.9,2.3,2.8)$ for 
$\alphabar(\rb) = (0.2,0.3,0.4)$ and $N_f = 2$.

Next, the hyperfine structure.  
The spin of each of the constituent quirks generates a dipole quirkcolor 
``magnetic'' field which leads to a spin-spin interaction,
\begin{eqnarray}
H_{hf} &=& \frac{8\pi}{3} \frac{\alphabar}{m_U m_D}
           {\vec S}_1\cdot {\vec S}_2 \, \delta^3({\vec r})+\ldots \; ,
\end{eqnarray}
where the terms we have neglected do not contribute to the 
angular momentum $\ell=0$ states.
This is the leading non-relativistic contribution to the
hyperfine structure.  (Relativistic corrections, such as Thomas
precession \cite{Brambilla:2004wf}, are small so long as the 
quirk masses are much larger than the strong scale.)
Sandwiching this Hamiltonian between states of
the unperturbed potential gives a splitting proportional to the
(unperturbed) wavefunction at the origin,
\begin{eqnarray}
\left|\psi_{1,0}(0)\right|^2 &=& 
     c_{1,0}\frac{ \left[\mu\, \alphabar(\rb) \right]^3}{\pi} \; .
\label{psi-0}
\end{eqnarray}
This is the familiar result from the hydrogen atom, except for the
constant $c_{n,\ell}$ which differs from one due to the $\log r$ 
term in our non-relativistic potential.
Numerically calculating the coefficient 
for $(n,\ell) = (1,0)$, we find 
$c_{1,0} \simeq (0.5,0.4,0.3)$ for $\alphabar(\rb) = (0.2,0.3,0.4)$
and $N_f = 2$.  The hyperfine splitting is thus estimated from 
Eq.~\eqref{psi-0} to be 
\begin{eqnarray}
E_{hf} &=&   c_{1,0} \frac{\mu^3\, \alphabar^4} {3 m_U m_D}
     \left\{ \begin{array}{rcl}
                 2 & \quad & \text{spin-triplet} \\ 
                -6 & \quad & \text{spin-singlet.} 
             \end{array} \right .
\label{hyper-eq}
\end{eqnarray}
As long as the quirk masses satisfy $|m_U-m_D| < E_{hf}$, 
the electrically neutral spin-singlet baryon $B^0_0$ is lighter 
than the electrically charged $q = (+1,0,-1)$ spin-triplet baryons 
$B^{q}_1$, in agreement with \cite{Chivukula:1989qb}.  
This requires our quirk current masses to be very nearly degenerate,
$m_U \simeq m_D$.
Hereafter, we use $B^0_0$ to denote our quirky dark matter candidate
in its ground state, $n=1$. 
We illustrate the spectrum of the ground and first excited baryonic 
states in Fig.~\ref{spectrum-fig}.

\section{Quirky Dark Matter Abundance}
\setcounter{equation}{0}

Stable quirks transforming under a chiral representation of the 
electroweak group have an abundance that is necessarily related 
to the baryon and lepton abundance through the electroweak phase
transition.  That such a relationship is inevitable was 
discovered in early work on the technibaryon abundance from 
technicolor theories \cite{Barr:1990ca}.  There it was shown that baryons
and technibaryons could have a common origin, since 
sphalerons intermix baryon, lepton, and technibaryon numbers.  
More interestingly, Ref.~\cite{Barr:1990ca} showed that the 
large mass of the technibaryons causes an additional Boltzmann 
suppression of their abundance, roughly scaling as $\exp[-M_*/T_*]$ where
$M_*$ is the mass of the technibaryon at the critical temperature
$T_*$ where sphalerons shut off.  This allows for TeV mass
technibaryons to nevertheless yield roughly the right dark 
matter abundance today (for recent calculations in technicolor
theories, see e.g.\ \cite{Gudnason:2006yj}).
There are three global flavor quantum numbers -- baryon, lepton, 
and technibaryon number -- while the sphaleron violates 
only one linear combination, leaving two anomaly-free 
invariants \cite{Kaplan:1991ah}.
Hence, while baryon and technibaryon numbers are related to
one another, one cannot determine technibaryon number
uniquely from just baryon number.  Instead, baryon, lepton,
and technibaryon numbers are ultimately determined in terms of 
linear combinations of two invariants which can be taken to be 
$B-L$ number \cite{Harvey:1990qw} and another combination involving 
both baryon (or lepton) number and technibaryon number
\cite{Kaplan:1991ah}.

The abundance of quirky dark matter is determined by an analysis
similar to that of technibaryon dark matter.  The main difference 
between our calculation below and that of \cite{Barr:1990ca} 
is that quirkcolor is assumed to be weakly-coupled through
the electroweak phase transition.  Sphalerons therefore yield 
an (asymmetric) abundance of quirks instead of quirky baryons.  To also 
exploit the Boltzmann suppression of quirky baryon number, quirks 
must acquire mass before sphalerons shut off, which can occur if the 
electroweak phase transition is not first order.  The Boltzmann suppression
for the abundance of quirks is therefore proportional to 
$\exp[-\lambda_q v(T_*)/T_*]$, where $v(T_*)$ is the electroweak vev 
at the critical temperature $T_*$.  The final ratio of 
quirky dark matter abundance to baryon abundance is determined 
by three quantities:  the two primordial anomaly-free $U(1)$ invariants 
and the ratio $m_q(T_*)/T_* = \lambda_q v(T_*)/T_*$.  In principle,
$v(T_*)$ and $T_*$ can be calculated within our theory.
This requires incorporating the effects of quirks on the phase
transition\footnote{Examples of theories with additional chiral fermions 
with large Yukawa couplings have been considered, e.g., \cite{Fok:2008yg}.
There it was found that the electroweak phase transition was 
\emph{weakened} (without superpartners), which is not inconsistent 
with our expectations,
though we leave a more detailed analysis to future work.}.

Following the classic calculation of \cite{Klinkhamer:1984di},
the divergence of the baryon, lepton, and quirky baryon currents 
can be constructed from
\begin{eqnarray}
\partial_\mu j^\mu = \frac{N g^2}{64 \pi^2} \epsilon^{\mu\nu\rho\sigma}
F^a_{\mu\nu} F^a_{\rho\sigma}
\end{eqnarray}
where only $SU(2)_W$ effects on $N$ electroweak doublets are considered. 
It is straightforward to determine that the sphaleron carries 
$B = N_g/2$, $L = N_g/2$ and $D = N_D/2$ charge
where $N_g = 3$ is the number of SM generations 
and $N_D = 1$ is the number of electroweak
doublets that carry quirky baryon charge.  We normalize the quirks
to carry $1/N_Q = 1/2$ quirky baryon charge, given $N_Q = 2$ quirkcolors,
precisely analogous to the $1/N_c = 1/3$ baryon number normalization 
of quarks.

\begin{figure}[t]
\includegraphics[width=0.44\textwidth]{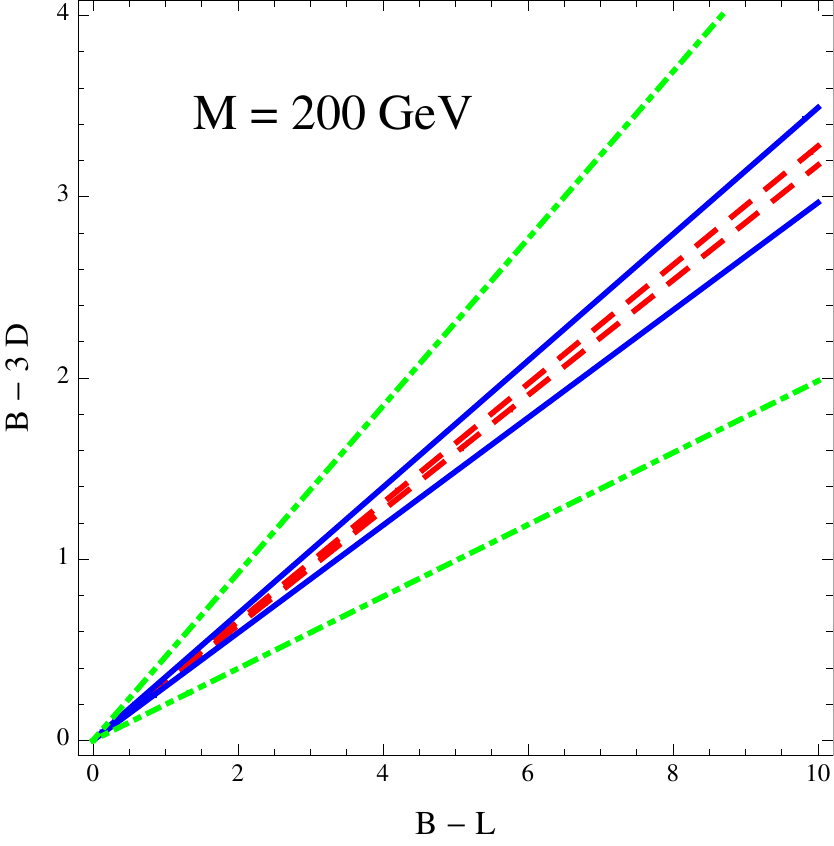}
\hfill
\includegraphics[width=0.44\textwidth]{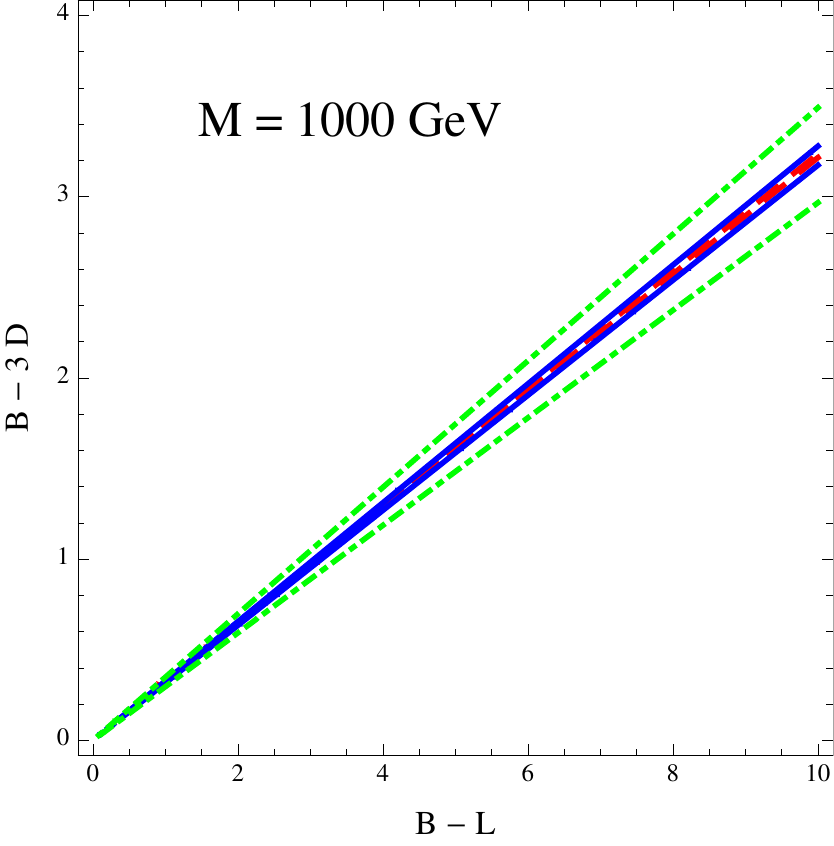} 
\caption{Contour plots of the density ratio 
$\rho_D/\rho_B = (1,5,25)$ shown by dashed, solid, dot-dashed
(red, blue, green) lines.
The axes are the invariants $(I_1,I_2) \equiv (B-L,B-3 D)$
in arbitrary units; a mirror symmetric plot can be obtained
taking $(I_1,I_2) \ra (-I_1,-I_2)$.
Plot on the left has $M = 200$~GeV, $x = 0.25$, and on the right
$M = 1000$~GeV, $x = 0.25$.  The plots demonstrate that a viable
region exists with $\rho_D/\rho_B \simeq 5$, corresponding to a
``bathtub ring'' around a valley in $(I_1,I_2)$ space.
The bottom of the valley has $\rho_D/\rho_B \simeq 0$.}
\label{density-fig}
\end{figure}

This result implies sphalerons violate the global $U(1)$ number 
$B + L + \frac{N_D}{N_g} D$.  The orthogonal combinations 
that are preserved can be taken to be $I_1 \equiv B - L$ and 
$I_2 \equiv B - \frac{N_g}{N_D} D$ (or $L - \frac{N_g}{N_D} D$) 
\cite{Kaplan:1991ah}.  Using the standard techniques
\cite{Harvey:1990qw,Dolgov:1991fr}, we enforce the following
relations among the chemical potentials:  (i) electric neutrality, and 
(ii) the vanishing of the chemical potential of the 
Higgs boson.  With these conditions, and taking $N_g = 3$ and 
$N_D = 1$, we obtain
\begin{eqnarray}
B &=& \frac{[36 f(x) + 4 f(x)^2] I_1 + [17 + 2 f(x)] I_2}{17 + 113 f(x) + 13 f(x)^2} \\
L &=& \frac{-[17 + 77 f(x) + 9 f(x)^2] I_1 + [17 + 2 f(x)] I_2}{17 + 113 f(x) + 13 f(x)^2} \\
D &=& f(x) \frac{[36 + 4 f(x)] I_1 - [111 + 13 f(x)] I_2}{51 + 339 f(x) + 39 f(x)^2}
\end{eqnarray}
where 
\begin{eqnarray}
f(x) &=& \frac{3}{2 \pi^2} \int_0^\infty 
\frac{y^2}{\cosh^2 \frac{1}{2} \sqrt{y^2 + x^2}}
\end{eqnarray}
in terms of $x = m_q(T_*)/T_*$.
The mass density ratio is, therefore, 
\begin{eqnarray}
\frac{\rho_D}{\rho_B} &=& \frac{D}{B} \frac{M}{m_p} \; = \; 
f(x) \frac{[36 + 4 f(x)] I_1 - [111 + 13 f(x)] I_2}{6 f(x) [18 + 2 f(x)] I_1 + [51 + 6 f(x)] I_2} 
\frac{M}{m_p} \; ,
\end{eqnarray}
where $M$ is the mass of $B^0_0$ in its ground state.
In Fig.~\ref{density-fig} we show contours of the resulting 
quirky dark matter density $\rho_D/\rho_B$ within the parameter space 
of the two primordial invariants $I_1$ and $I_2$.
Pure leptogenesis, which corresponds to $I_1 = -L$ and $I_2 = 0$, 
immediately implies $D = B/3$, independent of $f(x)$.  This mechanism 
is not viable since the mass of quirky dark matter would have to 
be $M \simeq 5 m_p B/D \simeq 15$ GeV, which is ruled out by direct collider
searches.  Pure baryogenesis or some mixture
of all three remains perfectly viable so long as the lepton number
of our universe remains unknown.

\section{Prospects for Direct Detection}
\label{sec:effective}
\setcounter{equation}{0}

\subsection{Overview}

There are three basic ways that quirky dark matter could 
potentially be found in direct detection experiments:
\begin{itemize}
\item[(i)] Elastic scattering through Higgs exchange.
\item[(ii)] Elastic scattering through higher dimensional 
operators.
\item[(iii)] Inelastic scattering through an excited quirky
baryonic state.
\end{itemize}

The third way, inelastic scattering, has been considered before
in general \cite{Pospelov:2008qx}
and recently in the context of 
composite inelastic dark matter \cite{Alves:2009nf,Kaplan:2009de}.
Quirky dark matter is more akin to Ref.~\cite{Pospelov:2008qx}, 
where it was shown that one needs fairly small splittings, up to about 
$10$~MeV, to allow for inelastic recombination.
The smallest splitting in quirky dark matter is the hyperfine splitting.  
Combining a rough bound from LEP II, $m_q \gsim 100$ GeV, 
with $\alphabar \gsim 0.1$ to satisfy direct detection bounds 
(explicitly shown later in this section), we find the hyperfine 
splitting $E_{hf} \simeq 2 \alphabar^4 \mu/3 \gsim 30$ MeV\@.
So, we do not anticipate inelastic scattering or inelastic 
recombination in direct detection experiments.

\subsection{Higgs Exchange}

Our quirks acquire mass through the Higgs mechanism, and hence
$B^0_0$ has interactions with matter
through Higgs exchange.
Just as the Higgs couples to the $\bar{q} q$ content of the
nucleon through $\langle N | \bar{q} q | N \rangle$, 
the Higgs also couples to the quirk-quirk content of our
quirky baryonic dark matter 
$\langle B^0_0 | q q | B^0_0 \rangle$.
Unlike the nucleon, however, the quirkcolor gluon condensate
is presumed to give a negligible contribution to the quirky
baryon mass.  The calculation of Higgs exchange is most easily
done in the low energy effective theory below the scale of
the quirky baryon.  Then we can treat $B^0_0$ as simply a
complex scalar with a renormalizable interaction with the Higgs,
\begin{eqnarray}
{\cal L} &\simeq& M \, h {B^0_0}^* {B^0_0}
\label{eff-scalar-int}
\end{eqnarray}
where this interaction assumes the mass of $B^0_0$ arises mostly
from the current quirk masses, i.e., $M \simeq m_U + m_D$.
With this interaction, we can use the results of
Ref.~\cite{Burgess:2000yq,Andreas:2008xy}, which considered 
the scattering of real scalars through Higgs exchange, and read off the 
direct detection cross section\footnote{Note that their 
$h S^2$ has a coefficient of $\lambda v_{246}$ which translates
into a Feynman rule with coefficient $2 \lambda v_{246}$ to
account for identical particles.  For a complex scalar, 
the equivalent Feynman rule constructed from Eq.(\ref{eff-scalar-int})
has no factor of $2$, 
and so our $\sigma({\rm nucleon})$ is smaller by a factor of $1/4$.}.  
We obtain
\begin{eqnarray}
\sigma({\rm nucleon}) &=& \frac{\mu(D,n)^2}{4 \pi A^2 m_h^4}
(Z f_p + (A-Z) f_n)^2
\end{eqnarray}
where $\mu(D,n)$ is the reduced mass of the $B^0_0$ and nucleon
for scattering off a nucleus with atomic number $Z$ and 
nucleon number $A$.
The nucleon parameters can be written as
\begin{eqnarray}
f_{\rm nucleon} &=& \frac{m_{\rm nucleon}}{v_{246}} \left[ 
\sum_{q=u,d,s} f^{({\rm nucleon})}_{Tq} + \frac{6}{27} f^{({\rm nucleon})}_{Tg}
                                  \right]
\end{eqnarray}
We have factored out the Higgs coupling to $B^0_0$,
so that only nuclear physics-dependent coefficients are 
present.  Numerically \cite{Ellis:2000ds},
\begin{eqnarray}
f^{(p)}_{Tu} \; = \; 0.020 & \qquad &  
f^{(p)}_{Td} \; = \; 0.026 \\
f^{(n)}_{Tu} \; = \; 0.014 & \qquad &  
f^{(n)}_{Td} \; = \; 0.036 
\end{eqnarray}
and \cite{Shifman:1978zn}
\begin{eqnarray}
f^{(p,n)}_{Tg} &=& 1 - \sum_{q=u,d,s} f^{(p,n)}_{Tq} \; .
\end{eqnarray}
The strange quark contribution to the nucleon is much more
uncertain.  A recent lattice calculation suggests it is much
smaller than has been previously estimated \cite{Ohki:2008ff}
(see also \cite{Giedt:2009mr}),
from which we extract $f^{(p,n)}_{Ts} \simeq 0.02$.

Interestingly, since $\mu(D,n) \simeq m_{\rm nucleon}$, there is essentially
no dependence of the nucleon cross section on the mass of the
dark matter.  This occurs because the (mass)$^2$ cancels out between
the numerator (its coupling to the Higgs squared) and denominator
(from the non-relativistic expansion of the cross section).
Putting all of this together, we obtain
\begin{eqnarray}
\sigma({\rm nucleon}) &\simeq& 1.8 \times 10^{-43} \left( \frac{114 \; {\rm GeV}}{m_h} \right)^4 \;\; {\rm cm}^2 \; .
\end{eqnarray}
The current best bounds come from CDMS \cite{Ahmed:2008eu}, 
$\sigma({\rm nucleon}) < 0.8$-$3.5 \times 10^{-43}$~cm$^{2}$, 
for dark matter mass between about $200$-$1000$~GeV\@.
This means that if the Higgs is
very near the LEP bound, quirky dark matter should be seen in
direct detection experiments in the very near future.
On the other hand, the absence of a direct detection signal
would put a lower bound on the Higgs mass that is consistent
with quirky dark matter.

\subsection{Higher Dimensional Operators}

The interaction of quirky dark matter with nuclei in direct
detection experiments can also proceed through higher dimensional 
operators involving the photon.
Since $B^0_0$ is an electrically neutral scalar composite, 
all its moments vanish.  The leading interactions are 
then the charge radius and the polarizability 
operators \cite{Pospelov:2000bq},
\begin{eqnarray}
L_\text{elastic}^\text{EM} &=& 
  \frac{1}{6} e r_D^2 \frac{\partial}{\partial x_i} E_i 
+ \frac{1}{2} \alpha_\text{pol} E^2 \; , 
  \label{H-em-elastic}
\end{eqnarray}
where $r_D$ is the charge radius, and $\alpha_\text{pol}$ is the
electromagnetic polarizability of $B^0_0$.
These interactions do not scale with the mass ($A$ number)
of the nucleus, and so we cannot use the usual 
$\sigma({\rm nucleon})$ cross section to compare with
experimental results.  Instead, we derive the {\it nuclear} 
cross sections that result from the charge radius
and polarizability.  These nuclear cross sections are 
in principle easy to compare with experiments, except that
experiments often quote bounds only on the average {\it nucleon} 
cross section.  Assuming the mass of a detector is dominated by
one (heavy) isotope of a nucleus with atomic number $A$, 
the relationship between the nucleon and the nucleus elastic 
scattering cross sections are related by 
\begin{eqnarray}
  \sigma({\rm Nucleus}) &=& \frac{\mu(D,N)^2}{\mu(D,n)^2} A^2 
      \, \sigma({\rm nucleon}) \; ,
  \label{eq:nucleon-sigma}
\end{eqnarray}
where $\mu(D,n)$ and $\mu(D,N)$ are the reduced mass of the 
dark matter with the nucleon and the nucleus respectively.

\subsection{Charge Radius}

The leading order interaction of a photon to a neutral scalar bound 
state of charged constituents is given by 
the charge radius. 
The scattering cross section off a \emph{nucleus} due to its charge
radius is given by \cite{Pospelov:2000bq}
\begin{eqnarray}
\sigma({\rm Nucleus})_{\text{charge radius}} &=& 
    \frac{16 \pi}{9} \mu(D,N)^2 \alphaem^2 Z^2 r_D^4
\label{Cscatter}
\end{eqnarray}
To gain a feeling for the size of the existing constraint, for example
from CDMS, we can compute the bound on $r_D^{2}$ for a few specific 
choices of dark matter mass:
\begin{eqnarray}
  r_D^{2} &\lsim& 
\left\{ \begin{array}{rcl}
        ( 510 \; \text{GeV})^{-2} & \quad & M = 200 \; {\rm GeV} \\
        ( 440 \; \text{GeV})^{-2} & \quad & M = 400 \; {\rm GeV} \\
        ( 370 \; \text{GeV})^{-2} & \quad & M = 800 \; {\rm GeV} \; . \\
        \end{array} \right.
\label{rD-bound}
\end{eqnarray}
We now compute the charge radius of $B^0_0$ in terms of the 
mass difference $\delta m_q = \left(m_U - m_D\right)$ 
and average mass $m_q = (m_U + m_D)/2$ of the quirks.
The magnitude of the charge radius is estimated 
by assuming the charge distribution inside the bound state
takes the form
\begin{eqnarray}
  \rho(r) & = & q e \left( \rho_{U}(r) - \rho_{D}(r) \right)  \nonumber \\
     & \approx & q e \left[
     \sum_i \left| \sqrt{2} \left( m_i \alphabar \right)^{3/2}
        \exp\left(-  m_i \alphabar r \right) \right|^2 \right] \; ,
\end{eqnarray}
where $\rho_i = \left| \langle B^0_0 | \psi_i\rangle \right|^2$ is the
probability density of finding the $i$-th quirk in the bound state
$B^0_0$.  We approximated individual quirk wavefunctions by assuming
the other quirk to be fixed in space and normalized it such a way that the 
total probability of finding the quirk is $1/2$.
The charge radius, interpreted as a measure of the size of
the bound state when probed electromagnetically, is defined classically
as the second moment of the spatial charge distribution,
\begin{eqnarray}
r_D^2 &=& \frac{1}{e} \int \! d^3 r \ r^2 \rho(r) \; = \; 
          \frac{3 q}{m_q^2 \alphabar^2}  \frac{\delta m_q}{m_q} +
          \mathcal{O}\left(\frac{\delta m_q}{m_q}\right)^2 \; , 
\end{eqnarray}
where we have assumed $\delta m_q \ll m_q$.  Using the constraint
from Eq.~\eqref{rD-bound}, we find
\begin{eqnarray}
   \frac{\delta m_q}{m_q} & \leq & 
       \left( \frac{\rb^{-1}}{25~\text{GeV}}  \right)^2 \times
\left\{ \begin{array}{rcl}
        6.4 \times 10^{-3} & \quad & M = 200 \; {\rm GeV} \\
        8.6 \times 10^{-3} & \quad & M = 400 \; {\rm GeV} \\
        1.2 \times 10^{-2} & \quad & M = 800 \; {\rm GeV} \; .
        \end{array} \right.
\end{eqnarray}
Hence, the quirk masses must be very nearly degenerate to avoid
generating an electromagnetic charge radius that exceeds the
direct detection bounds.  Interestingly, we already required 
$\delta m_q < E_{hf}$, to ensure the lightest quirky baryon
is electrically-neutral.  We see that a self-consistent picture
has emerged in which nearly or exactly degenerate quirks ensures
both that the lightest quirky baryon is electrically neutral
as well as a negligible electromagnetic charge radius.
In Appendix~\ref{UD-parity-app}, we demonstrate that the 
vanishing of the charge radius can result from imposing an exact 
discrete symmetry, ``UD-parity'', which enforces $m_U = m_D$.

\subsection{Polarizability}

Having discussed and discarded the charge radius operator, 
we now move on discuss the electromagnetic polarizability.
The scattering cross section due to the polarizability operator 
is given by \cite{Pospelov:2000bq}
\begin{eqnarray}
\sigma({\rm Nucleus})_{\text{pol}} &\approx& 
  \frac{144}{25} \, \mu(D,N)^2 Z^4 \alphaem^2 \frac{\alpha_{\rm pol}^2}{r_0^2}
  \; ,
\label{eq:sigma-pol}
\end{eqnarray}
where the nucleus is assumed to be a sphere of homogeneous charge
with radius $r_0 = \sqrt[3]{A}\times 1.2~\text{fm}$. 
To again gain a feeling for the constraint, for example from CDMS, 
we compute the bound on $\alpha_\text{pol}$ for a few specific 
choices of dark matter mass,
\begin{eqnarray}
  \alpha_\text{pol} & \lesssim & 
\left\{  \begin{array}{rcl}
         \left( 106~\text{GeV}\right)^{-3} & \quad & M = 200 \; {\rm GeV} \\
         \left(  95~\text{GeV}\right)^{-3} & \quad & M = 400 \; {\rm GeV} \\
         \left(  85~\text{GeV}\right)^{-3} & \quad & M = 800 \; {\rm GeV} \; .
         \end{array} \right. 
\label{pol-bound}
\end{eqnarray}

We now calculate the polarizability of $B^0_0$.  To proceed, we first
quickly review the standard polarizability calculation.  If an electric 
field $\mathcal{E}$ is applied to the bound state in the $z$ direction, 
it causes a perturbation to the Hamiltonian
\begin{eqnarray}
  H_\text{pert} &=& q e \mathcal{E} z 
\end{eqnarray}
where $q e$ is the constituent quirk charge.  The Schr\"odinger equation 
in the presence of this perturbation,
\begin{eqnarray}
  \left( H_0 +  H_\text{pert} \right) |\psi\rangle &=& E |\psi\rangle 
\end{eqnarray}
can be approximately solved by standard perturbation theory
methods,
\begin{eqnarray}
  |\psi\rangle &=& |0\rangle + q e \mathcal{E} \sum_{k>0} 
                   \frac{\langle k| z | 0 \rangle}{E_0 - E_k} |k\rangle \; ,
\end{eqnarray}
where $H_0 |k\rangle = E_k |k\rangle$, and the ground state is
$|0\rangle$.  In a moment we will identify the states $\{ |k\rangle \}$
with the relevant states of $B^0_0$. 
At leading order, the dipole moment of this state in
the $z$ direction is composed of two terms:
\begin{eqnarray}
  p_z &=&  - q e  \langle \psi | z | \psi \rangle  \nonumber \\
      &=&  - q e  
           \left\{ \langle 0| z | 0 \rangle + \sum_{k>0} \left[ 
               \frac{\langle k| q e \mathcal{E} z | 0 \rangle}{E_0 - E_k}  
               \langle 0| z | k \rangle 
             + \frac{\langle 0| q e \mathcal{E} z | k \rangle}{E_0 - E_k} 
               \langle k| z | 0 \rangle \right] 
           \right\}
\end{eqnarray}
The first term, $\langle 0|z|0\rangle$, is the moment of the 
unperturbed state (if any). 
The second set of terms represent the moment induced by 
the electric field, $p_\text{ind} = \alpha_\text{pol} \mathcal{E}$, 
where $\alpha_\text{pol}$ is defined as the polarizability.
We thus find
\begin{eqnarray}
  \alpha_\text{pol} &=& 2 q^2 e^2 \sum_{k>0}  
  \frac{ \left| \langle k|z|0 \rangle \right|^2}{E_0 - E_k} \; .
\label{def-pol}
\end{eqnarray}

In order to extract the proper dependence of the matrix element 
in Eq.~\eqref{def-pol} on $\alphaem$ and $\alphabar$, we must also 
include the ordinary electromagnetic Coulomb potential.
The unperturbed potential is then simply
$V(r) = - \left( \alphabar(r) + q^2 \alphaem \right)/r$, 
the effective Bohr radius then is given by 
$\rb = [(\alphabar(\rb) + q^2 \alphaem) \mu]^{-1}$ 
and the energy eigenvalues are given as 
$E_k = - \left( \alphabar(\rb) + q^2 \alphaem \right)^2\mu/2k^2$.  
Denoting the states by the usual quantum 
numbers,  
$\{ |k \rangle \} = \{| n, l, m \rangle \}$, 
with the ground state $|0 \rangle =  | 1, 0, 0 \rangle$,
we find
\begin{eqnarray}
  \alpha_\text{pol} &=& 2 q^2 e^2 \sum_{n>1}  \frac{ \bigl| 
       \langle n, 1, 0| z |1,0, 0 \rangle \bigr|^2} {E_1 - E_n} 
       \nonumber \\ 
& = & k_{pol} \, \frac{9}{2} \,  
      \frac{q^2 \alphaem }{\alphabar + q^2 \alphaem} \: \rb^3 
\label{expn-pol}
\end{eqnarray}
The result above shows that there is an additional 
$q^2\alphaem/(\alphabar + q^2 \alphaem)$
suppression in the polarizability relative to the hydrogen atom.  
This agrees with the analogous calculation for the electromagnetic 
polarizability of heavy quarkonia, substituting the quirkcolor 
coupling with the QCD coupling \cite{Chen:1997zza}.
This factor arises since the binding potential is proportional to 
$\alphabar + q^2 \alphaem$.  The additional non-Abelian correction 
resulting from the $\log r$ term in the potential is encoded
in the coefficient $k_{pol}$.  Numerically solving the Schr\"odinger
equation, we calculated the corrections to the first few terms,
finding the largest correction to the $\langle 2,1,0|z|1,0,0\rangle$ term
and a smaller correction to $\langle 3,1,0|z|1,0,0\rangle$ term.
Extrapolating from these results, our numerical estimate is
$k_{pol} = (1.3,1.4,1.5)$ for $\alphabar(\rb) = (0.2,0.3,0.4)$ 
and $N_f = 2$.

\begin{figure}[t]
\centering
\includegraphics[totalheight=0.35\textheight]{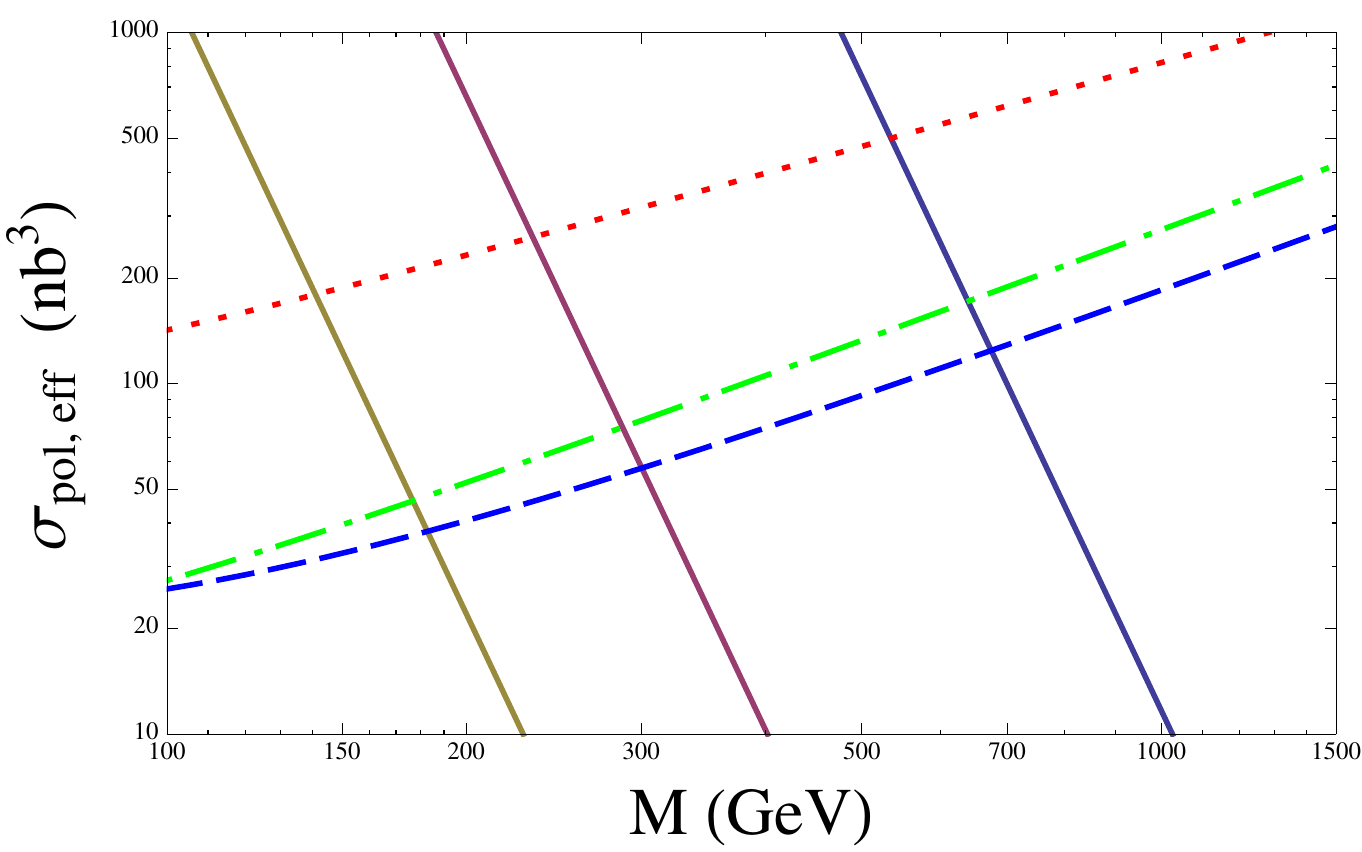} 
\caption{Bounds of nuclear cross section. We have plotted 
$\sigma_\text{pol,eff}$ from Eq.~\eqref{sigma-eff} using data from 
CDMS (blue dash), Xenon (green dot-dash) and Tungsten (red dot).  
The solid lines are theory predictions due to the polarizability operator 
for two flavors as a function of dark matter mass $M_D$. 
From right to left we have plotted $\sigma_\text{eff}$ for 
$\alphabar(\rb) = 
\{0.2(\text{blue}), 0.4(\text{purple}), 0.6(\text{yellow}) \}$
respectively.}
\label{polarizabilitybounds-fig}
\end{figure}

We can now easily determine the bounds and prospects for detection of
quirky dark matter as a function of $\alphabar$ using our 
expression for $\alpha_\text{pol}$ in Eq.~\eqref{expn-pol},
In Fig.~\ref{polarizabilitybounds-fig} we show theory predictions 
and experimental bounds on an effective cross section. 
To combine results from several experiments with theory predictions 
the effective nuclear cross section applicable to the polarizability
operator is 
\begin{eqnarray}
\label{sigma-eff}
  \sigma_\text{pol,eff} &=& 
  \frac{r_0^2}{Z^4} \, \frac{1}{\mu(D,N)^2} \sigma({\rm Nucleus})_{\text{pol}} 
  \; .
\end{eqnarray}
where the individual experiments' target size ($r_0$) and charge
($Z$ value) have been factored out, along with two powers of
the reduced mass, analogous to Eq.~(\ref{eq:nucleon-sigma}).

To gain an appreciation of the constraint
on $\rb^{-1}$, we can compute the lower bounds on the 
inverse Bohr radius derived from the polarizability constraint 
directly from Eqs.~(\ref{expn-pol}),(\ref{pol-bound}):
\begin{eqnarray}
\rb^{-1} &\gsim& 
    \left\{ \begin{array}{rcl}
            40-32 \; {\rm GeV} &\quad& M = 200 \; {\rm GeV} \\
            33-27 \; {\rm GeV} &\quad& M = 400 \; {\rm GeV} \\
            30-24 \; {\rm GeV} &\quad& M = 800 \; {\rm GeV} \; ,
\end{array} \right.
\end{eqnarray}
where the range in value corresponds to $0.2 < \alphabar < 0.6$.
Since the cross section induced by the electromagnetic 
polarizability is proportional to $\rb^{6}$, there is
a relatively minor dependence on the mass of dark matter
and the strength of the strong force.

\section{Indirect Detection through Absorption Lines}
\label{sec:absorption}
\setcounter{equation}{0}

An interesting feature of quirky dark matter is that it has
a spectrum of excited bound states that can be excited by absorption
of SM particles such as the photon and $Z$.  The photon
interaction, in particular, provides an fascinating possibility 
to probe dark matter \emph{directly} through photon absorption lines,
entirely analogous to how matter itself is probed through its own 
photon absorption lines.  The typical energies for quirky excitations, 
as we see below, are in the gamma-ray region.  The possible existence
of dark lines due to inelastic transitions of dark matter was considered 
before in a somewhat different setup in Ref.~\cite{Profumo:2006im}.
We will apply their results, suitably modified for our case 
(with one correction to their formulae given in
Appendix \ref{PS-app}), to the two transitions of greatest interest:
the quirky Lyman-alpha transition, and the quirky hyperfine transition,
illustrated in Fig.~\ref{spectrum-fig}.

\subsection{Quirky Lyman-alpha}

The photon energy corresponding to quirky Lyman-alpha transition,
between the ground state and the first excited state 
can be read off from Eq.~(\ref{Lyman-eq}),
\begin{eqnarray}
E_{L\alpha} &=& E_2-E_1 = \frac{4 k_1 - k_2}{3} \, 
\frac{8}{3} \alphabar^2 \mu \; .
\end{eqnarray}
The deviation from purely Coulombic is characterized by the constant 
$(4 k_1 - k_2)/3 \simeq 
(1.0,1.0,0.9)$ for $\alphabar(\rb) = (0.2,0.3,0.4)$,
which is negligible for the quirky Lyman-alpha transition.

The width of this absorption feature can be computed in 
the Coulombic approximation by analogy with hydrogen Lyman-alpha.  
We obtain
\begin{eqnarray}
\Gamma_{L\alpha} &=& \frac{4}{9} q^2 \alphaem E^3_{L\alpha} 
|\langle 1|\bar{r}|0\rangle|^2
\; = \; \frac{64}{6561} \alphaem \alphabar^4 \mu
\end{eqnarray}
This width, taking $\alphabar \rightarrow \alphaem$ 
and $q \rightarrow 1$ reproduces the well known value for 
hydrogen Lyman-alpha.  
Plugging in representative values, the width is
\begin{eqnarray}
\Gamma_{L\alpha} &=& 5 \times 10^{-4} \; {\rm GeV} 
           \left( \frac{\alphabar}{0.4} \right)^3
           \left( \frac{\rb^{-1}}{25 \; {\rm GeV}} \right) \; .
\end{eqnarray}
It will also prove convenient to express the width as 
\begin{eqnarray}
\frac{\Gamma_{L\alpha}}{E_{L\alpha}} &=& 
\frac{512}{19683} \alphaem \alphabar^2 \; \simeq \; 
1.9 \times 10^{-4} \, \alphabar^2 
\label{funnyLalphawidth}
\end{eqnarray}

\subsection{Quirky Hyperfine}

The next transition of interest arises from the quirky hyperfine structure.
The energy splitting between the $B^0_1$ and $B^0_0$ states is given by
\begin{eqnarray}
E_{hf} &=& c_{1,0} \frac{2}{3} \alphabar^4 \mu 
\end{eqnarray}
where $c_{1,0}$ is an order one constant that characterizes
the deviation from purely Coulombic.  A few specific values
were computed in Sec.~\ref{binding-sec}.

The decay rate of the quirky hyperfine transition can also be computed.
The result in the purely Coulombic approximation is the same as for 
quarkonia \cite{Brambilla:2005zw}
\begin{eqnarray}
\Gamma_{hf} = \frac{4}{3} \alphaem q^2 \frac{E_{hf}^3}{\mu^2}
\; = \; \frac{8}{81} \alphaem \alphabar^{12} \mu \; .
\end{eqnarray}
Plugging in representative values, we find the width to be 
\begin{eqnarray}
\Gamma_{hf} &=& 8 \times 10^{-7} \; {\rm GeV} 
           \left( \frac{\alphabar}{0.4} \right)^{11}
           \left( \frac{\rb^{-1}}{25 \; {\rm GeV}} \right)
\end{eqnarray}
It will again prove convenient to express the width as 
\begin{eqnarray}
\frac{\Gamma_{hf}}{E_{hf}} &=& 
\frac{4}{27} \alphaem \alphabar^8 \; \simeq \; 
1.1 \times 10^{-3} \, \alphabar^8 
\label{funnyhyperwidth}
\end{eqnarray}

\subsection{Absorption and Broadening}

For these transitions to be visible to gamma-ray observatories, 
three things are required:
(1) the cross section on resonance should be substantial, 
(2) the width of the dark matter Doppler-broadened resonance 
should be resolvable, and  
(3) the energy of the transition should be within the range
that is observable.  Several observatories view the universe 
in the relevant energy range, satisfying (3).  In this section we 
compute the cross section and Doppler-broadened width.

The absorption cross section near a resonance takes the usual 
Breit-Wigner form \cite{Weinberg:1995mt}
\begin{eqnarray}
\sigma_{\rm abs} &=& \frac{6 \pi}{p_{\rm cm}^2}
                   \frac{M_*^2 \Gamma_*^2}{(s - M_*^2)^2 + M_*^2 \Gamma_*^2},
\label{breit-wigner-eq}
\end{eqnarray}
where $M$ is the mass of the ground state (the mass of quirky dark matter),
and $M_* = M + E_\gamma$ is the mass of the excited state. 
The dominant decay of the resonance is into photons, 
$\Gamma_{*} \simeq \Gamma_{M_* \rightarrow M\gamma}$, since 
decay to an on-shell $Z$ is kinematically forbidden
(throughout our parameter space), while decay through a 
virtual photon to a fermion pair is suppressed by $\alphaem$.
The factor of $6 \pi$ comes from 
$4 \pi (2 j_R + 1)/((2 s_1 + 1) (2 s_2 + 1)$ where 
$j_R = 1$ is the massive spin-1 resonance, $(2 s_1 + 1) = 2$ for
the massless photon, and $s_2 = 0$ for the massive scalar quirky
dark matter particle.

There are two potential methods to exploit this absorption cross section.
The first, and most promising, is to look towards massive galaxy clusters 
that have the largest column density of dark matter and a large dark matter 
velocity dispersion.  We then compare this to the seemingly less promising
method of looking for absorption over cosmological distances. 

We follow the formalism of \cite{Profumo:2006im} to determine 
the effect of the broadening and the size of the absorption cross section.  
This formalism applies to any of the photon-induced baryonic excitations,
including quirky Lyman-alpha and quirky hyperfine.
The resonance energy is 
\begin{eqnarray}
\Eres &=& \frac{M_*^2 - M^2}{2 M}
\end{eqnarray}
which is roughly equal to the mass difference, $M_* - M$
for $\alphabar \lsim 1$.

The dark matter velocity distribution within a galaxy cluster is assumed 
to follow a Maxwell-Boltzmann velocity distribution, 
denoted $f_{\rm MB}(p)$.  The effective absorption cross section 
for a photon of energy $E_\gamma$ is
\begin{eqnarray}
\sigma(E_\gamma) &=& 
    \int_0^\infty dp \, f_{\rm MB}(p) \langle \sigma \rangle,
\end{eqnarray}
where
\begin{eqnarray}
\langle \sigma \rangle &=& 
   \int_{-1}^1 \frac{d \, \cos\theta}{2} \frac{6 \pi}{p_{\rm cm}^2} 
   \frac{M_*^2 \Gamma_*^2}{(s - M_*^2)^2 + M_*^2 \Gamma_*^2},
   \label{sigmu-eq}
\end{eqnarray}
is the total cross section after integrating over the 
incident angle.  The center of mass energy is
$s = M^2 + 2 E_\gamma\left( \sqrt{p^2 + M^2}- p \cos\theta \right)$
while the momentum in the center-of-mass frame is given by
\begin{eqnarray}
p_{\rm cm}^2 &=& \frac{(M^2 - s)^2}{4 s} \; .
\end{eqnarray}
The integral in Eq.~(\ref{sigmu-eq}) was solved analytically
in the Appendix of Ref.~\cite{Profumo:2006im}, which we checked 
and agree with except for a correction to one expression given in 
our Appendix \ref{PS-app}.

For our case, with $M_* \simeq M \gg E_\gamma$, the analytic formula can
be drastically simplified in the following limits.
First, observe that $M_* \Gamma_* \ll \Delta M^2$ which is 
equivalent to $2 \Gamma_* \ll \Eres$, is automatic for our 
perturbative calculation of the width of the resonance to be valid. 
Next, consider the limit 
\begin{eqnarray}
\Gamma_* M_* \ll 2 \Eres p
\end{eqnarray}
which corresponds to roughly
\begin{eqnarray}
\Gamma_* \ll 2 \Eres \sigma_v
\label{gammasigv-condition-eq}
\end{eqnarray}
given the average momentum of dark matter with a Maxwell-Boltzmann
distribution is roughly $p \simeq M \sigma_v$.  Now compare this
expression with Eqs.~(\ref{funnyLalphawidth}),(\ref{funnyhyperwidth}).
For $\alphabar \lsim 1$, Eq.~(\ref{gammasigv-condition-eq}) is satisfied 
for $\sigma_v \gsim 10^{-4}$, which is itself satisfied by 
large galaxy clusters.

Putting all this together, we obtain the following simple formula
for the cross section on resonance,
\begin{eqnarray}
\langle \sigma \rangle|_{\rm res} = 
   \frac{3}{2} \pi^2 \frac{\Gamma_* M}{\Eres^3 p}
\end{eqnarray}
which can be integrated over a Maxwell-Boltzmann distribution
to become
\begin{eqnarray}
\sigma(E_\gamma = \Eres) = 
   \frac{3}{\sqrt{2}} \pi^{3/2} \frac{\Gamma_*}{\Eres^3 \sigma_v} \; .
\end{eqnarray}

Interestingly, this form of the on-resonance cross section has
several important features:  First, there is no explicit dependence
on the mass of the particle.  Second, the resonance photon energy 
dependence in $\Gamma_*$ (from Lyman-alpha or hyperfine) exactly 
cancels the dependence in the denominator.  This leads to very 
simple expressions for the on-resonance absorption cross section
\begin{eqnarray}
\sigma(E_\gamma = \Eres) &=& 
   \frac{16384 \sqrt{2} \pi^{3/2}}{59049} \alphaem
   \frac{\rb^2}{\sigma_v}
   \qquad (L\alpha) \\
\sigma(E_\gamma = \Eres) &=& 
   \frac{\pi^{3/2}}{\sqrt{2}} \alphaem \alphabar^2
   \frac{\rb^2}{\sigma_v}
   \;\;\;\qquad\qquad (hf) 
\end{eqnarray}
The opacity to $\gamma$-rays due to these dark lines on resonance 
can be estimated using the optical depth $\tau = \sigma \Sigma_{\rm DM}/M$
where $\Sigma_{\rm DM}$ is the dark matter surface density
associated with the integral along the line of sight of the
dark matter density.  As an example, consider the Coma cluster,
which has a surface density that was estimated by Ref.~\cite{Profumo:2006im}
to be $\Sigma_{DM} \simeq 5 \times 10^{26} \; {\rm GeV/cm}^2$ 
with velocity dispersion $\sigma_v = 820 \; {\rm km/s}$.
Plugging in these characteristic values,
\begin{eqnarray}
\tau|_{\rm res} &=& 1 \times 10^{-5} 
\frac{820 \; {\rm km/s}}{\sigma_v}
\left( \frac{25 \; {\rm GeV}}{\rb^{-1}} \right)^2
\frac{200 \; {\rm GeV}}{M}
\frac{\Sigma_{\rm DM}}{5 \times 10^{26} \; {\rm GeV/cm}^2} 
\qquad (L\alpha) \quad \\
\tau|_{\rm res} &=& 1.5 \times 10^{-5} \alphabar^2
\frac{820 \; {\rm km/s}}{\sigma_v}
\left( \frac{25 \; {\rm GeV}}{\rb^{-1}} \right)^2
\frac{200 \; {\rm GeV}}{M}
\frac{\Sigma_{\rm DM}}{5 \times 10^{26} \; {\rm GeV/cm}^2} \quad (hf) \quad
\end{eqnarray}
we see that $\tau$ is much smaller than one.  
For the Coma cluster, only a small fraction of photons are
expected to be absorbed in either the Lyman-alpha or hyperfine
transitions.  

There are three potential ways to improve on this result.  
The first is to search for systems with larger surface mass densities.
This is most likely to arise in the largest clusters that are also
the most compact, and thus have a very concentrated mass function.
The second is to perform more precise measurements of the photon flux 
of particular clusters, which would allow probing optical depths 
considerably smaller than one.  Third, combining 
$\gamma$-ray spectra from many different clusters of different mass, 
velocity dispersion, and redshift, and suitably processing them into 
a normalizable spectrum, one could significantly improve the search
for dark lines through large scale galaxy cluster sampling.  
Given that the cluster number density
is rising rapidly as the cluster mass is decreased
(see, e.g., \cite{Warren:2005ey}), and with improved photon
flux and energy resolution, this  is probably the best approach for
the future.

Assuming methods are developed to effectively probe these 
small optical depths, it is also important to know the
Doppler-broadened resonance width.  In our case, the Doppler-broadened 
width arises from the dark matter velocity distribution in 
the observed systems.  For a Maxwell-Boltzmann velocity distribution, 
we find the observed width of the resonance is well-fit to an Gaussian 
\begin{eqnarray}
\sigma(E_\gamma) &=& \sigma(\Eres) 
    \exp \left[ -\frac{(E_\gamma - \Eres)^2}{2 (\sigma_v \Eres)^2} \right]
\end{eqnarray}
with a fractional width $\Delta E/E \simeq \sigma_v$.

The velocity dispersion of dark matter in galaxy clusters 
has been found to scale as \cite{Evrard:2007py}
\begin{eqnarray}
\sigma_v \simeq (1080 \; {\rm km/s}) 
\left( \frac{h(z) M_{\rm clus}}{10^{15} M_\solar} \right)^{0.336}
\end{eqnarray}
where $h(z) = H(z)/100 \; {\rm km} \, {\rm s}^{-1} \, {\rm Mpc}^{-1}$ 
is the normalized 
Hubble parameter at redshift $z$ and $M_{\rm clus}$ is the cluster mass
(defined as the mass within a sphere encompassing a mean mass
density of 200 times the background matter density at
that redshift).
Searching for dark lines of dark matter in clusters thus requires
the gamma-ray energy resolution comparable to the velocity dispersion. 
The full width at half maximum (FWHM) is approximately 
$2.35 \sigma_v \simeq 0.003 - 0.01$
for the largest galaxy clusters with mass between about 
$10^{14-15} M_\solar$.  Interestingly, for soft $\gamma$-rays
up to 8 MeV, this resolution was achieved with the 
INTEGRAL spectrometer \cite{Schanne:2006yn}.
The EGRET and Fermi/GLAST observatories extend up to much
higher gamma-ray energies, 30 GeV and about 1 TeV, respectively\@.  
Unfortunately, the FWHM energy resolution 
of these instruments is roughly $0.2$ for EGRET and between 
$0.05 - 0.1$ for Fermi/GLAST, which is a just bit too course to 
likely resolve the $\gamma$-ray dark line feature that we predict 
in our model.

Finally, we consider the effect of scattering over cosmological distances.
At a redshift of $z \sim 1$ or larger, a photon with a given 
initial energy $E_\gamma$ will sweep out a resonance of width 
$\Gamma \sim z E_\gamma$ (e.g.\ \cite{Goldberg:2005yw,Hooper:2007jr}).
In order for this to give an observable effect, however, the mean 
free path $\ell$ for photon absorption must be shorter than the 
cosmological distance that the photon spends on resonance
\begin{eqnarray}
\ell &=& \frac{M}{\rho_D \langle \sigma \rangle|_{\rm res}} \; \lesssim \; 
H^{-1} \frac{\Gamma}{E_\gamma} \; .
\end{eqnarray}
Dark matter in the cosmos has negligible momentum, and so the
resonance cross section can be obtained from Eq.~(\ref{breit-wigner-eq})
in the limit $p \ll M$, 
$\langle \sigma \rangle|_{\rm res} = 6 \pi / p^2_{\rm cm} = 6 \pi/E^2_\gamma$.
Substituting, we obtain
\begin{eqnarray}
\frac{\ell}{H^{-1}} &\simeq& 
    2 \times 10^4 \, 
    \frac{M}{200 \; {\rm GeV}} \, 
    \left( \frac{E_\gamma}{100 \; {\rm MeV}} \right)^2 \; ,
\end{eqnarray}
which shows that even if $\Gamma \sim E_\gamma$, the photon does 
not travel nearly far enough to be absorbed over a cosmological distance.

\section{Quirkcolor Glueball Decay}
\label{sec:glueballdecay}
\setcounter{equation}{0}

With exactly two flavors, our quirkcolor theory confines.
It is straightforward to estimate the confinement scale
$\Lambda_Q$.  Quirky theories suffer from a potential 
cosmological problem, namely photon injection
during nucleosynthesis, if glueballs decay into photons 
with a lifetime that is of order $1$~s.

Confinement gives mass to the quirkcolor ``glueballs'' that we
crudely approximate to have mass $\Lambda_Q$.  
For theories with vector-like quirks, 
Refs.~\cite{Kang:2008ea,Juknevich:2009ji} showed that the 
glueballs decay slowly, since the leading operators are 
suppressed by many powers of the quirk mass.  One such 
operator is
\begin{eqnarray}
\frac{q^2\alpha_{\rm em}\alphabar(m_q)}{m_q^4} 
F^{\mu\nu}_Q {F_Q}_{\mu\nu} F^{\rho\sigma} F_{\rho\sigma}
\label{glueball-decay-photon-eq}
\end{eqnarray}
where $F_Q$ and $F$ are the field-strengths of the 
quirkcolor group and electromagnetism, respectively.
This results in a decay rate 
\begin{eqnarray}
\Gamma &\sim& \sum_{\rm quirks} 
\frac{q^4\alpha_{\rm em}^2\alphabar(m_q)^2}{8 \pi} 
\frac{\Lambda_Q^9}{m_q^8} \; \simeq \;
\left( \frac{\Lambda_Q}{1 \; \mathrm{GeV}} \right)^9
\left( \frac{100 \; \mathrm{GeV}}{m_q} \right)^8
\left( \frac{\alphabar(m_q)}{0.1} \right)^2 \;\; \mathrm{s}^{-1} \; .
\end{eqnarray}
where the sum is over all of the quirks given in 
Table~\ref{table:model}.

A different operator exists in our model due to the 
Higgs coupling to our chiral quirks.  Integrating out 
quirks and the scalar Higgs boson simultaneously
results in a dimension-7 operator
\begin{eqnarray}
\frac{\alphabar(m_q) m_f}{4 \pi v^2 m_h^2} 
F^{\mu\nu}_Q {F_Q}_{\mu\nu} \bar{f} f
\label{glueball-decay-higgs-eq}
\end{eqnarray}
where $v = 174$ GeV\@. 
This leads to glueball decay into a pair of light SM fermions 
that satisfies $2 m_f \lsim \Lambda_Q$.
Despite the lower dimensionality, this operator is not obviously 
less suppressed than Eq.~(\ref{glueball-decay-photon-eq}), 
due to the Yukawa suppression $m_f/v$.  The decay rate is
\begin{equation}
\Gamma \; \sim \; \sum_{\rm quirks} 
\frac{\alphabar(m_q)^2 m_f^2}{8 \pi} \frac{\Lambda_Q^7}{v^4 m_h^4} 
\; \simeq \; 
\left( \frac{\Lambda_Q}{1 \; \mathrm{GeV}} \right)^7
\left( \frac{115 \; \mathrm{GeV}}{m_h} \right)^4
\left( \frac{\alphabar(m_q)}{0.1} \right)^2
\left( \frac{m_f}{0.1 \; \mathrm{GeV}} \right)^2 \;\; \mathrm{s}^{-1} \, .
\end{equation}
We see this decay rate is comparable to the rate into photons 
for the example parameters.
The main distinction we draw is that the Higgs-mediated decay
does not depend on the quirk mass.  Hence, we can contemplate 
quirk masses that significantly exceed $100$~GeV without 
necessarily leading to cosmological difficulties of late
decaying glueballs, so long as the Higgs is relatively light.
We emphasize that while our estimates are parametrically correct,
they nevertheless have significant uncertainties, 
particularly with regard to the identification of the 
glueball mass with $\Lambda_Q$.

\section{Discussion}
\label{sec:concl}
\setcounter{equation}{0}

We have presented a new theory of dark matter that is
made up of a baryonic composite of a quirks that transform
under a new strongly coupled sector, $SU(2)_Q$ quirkcolor.
The abundance of quirky dark matter is linked to the 
baryonic abundance through electroweak sphalerons.

The baryonic excitation spectrum was computed, including
the fine and hyperfine structure.  The lightest baryonic
state can be made automatically charged-neutral when
the quirks have (nearly) degenerate masses, which can
be ensured through a discrete symmetry (UD-parity).
Degenerate quirk masses also eliminates the dimension-6
electromagnetic charge radius operator, allowing a much
larger range of scales to be considered.

Quirky dark matter is at the threshold of direct direction
though elastic nuclear recoil.  Two processes lead to 
nuclear recoil cross sections through 
(i) Higgs exchange, which couples proportional
to the (mass)$^2$ of the nuclei (as usual), and 
(ii) electromagnetic polarizability, which couples to the 
electric charge $Z^4$.

Indirect detection may be possible by searching for
a gamma-ray ``dark line'' spectroscopic feature in 
galaxy clusters that results from the quirky Lyman-alpha or
quirky hyperfine transitions.  This is a difficult measurement
that might be possible in  the future.  It requires sensitivity to optical
depths much smaller than one.  We envision this could be 
accomplished with excellent gamma-ray spectral sensitivity
applied to a large number of galaxy clusters, suitably 
combining the results together.  The feature itself has a
Doppler-broadened FWHM roughly of order 
$2.35 \Delta E_\gamma/E_\gamma \simeq 2.35 \sigma_v \simeq 0.003 - 0.01$.
This is close to but somewhat smaller than the FWHM 
energy resolution of EGRET and Fermi.
Indirect detection through other means, such as annihilation
in the Sun, galaxy, or beyond, does not occur so long as 
the full global $U(1)_{\rm QB}$ quirky baryon number is exact.  
Annihilation signals would reappear if quirky baryon number 
were broken to a $Z_2$, and signals of dark matter decay would
result if $U(1)_{\rm QB}$ were completely broken 
(but only very, very slightly).  

The collider signals of quirky dark matter represent a plethora of 
possibilities \cite{Kang:2008ea}.
Quirks can be pair-produced, which form bound
states with quirkcolor strings attached.  They will stretch
and flop back and forth shedding angular momentum in some
combination of quirky glueballs (from the quirks and the quirkcolor
string) as well as photon emission from the quirks
which may result in interesting underlying event signals
\cite{Harnik:2008ax}.  Eventually the quirks bound in a mesonic state 
recombine and annihilate, somewhat analogous to heavy quarkonia
annihilation.  Quirky baryon production is kinematically suppressed 
due to the need to pair produce a baryon and anti-baryon
(four quirks total), to conserve quirky baryon number.
Clearly, the collider physics of quirky dark matter 
is an area ripe for future exploration.

\begin{appendix}
\refstepcounter{section}

\section*{Appendix~\thesection:~~Non-relativistic Details}
\label{NR-app}

The construction of the non-relativistic theory starting from the theory 
described in Sec.~\ref{field-content-sec} begins with the chiral
two component spinors in Table.~\ref{table:model}.
Below the electroweak scale, the two-component fermions can be written
in terms of 2 four-component Dirac fermions and their charge conjugates 
for the quirks and the anti-quirks respectively,
\begin{equation}
\begin{split}
f_U \equiv \begin{pmatrix} \xi_U \\ \xi_{\bar{U}}^\dag \end{pmatrix}, 
    \quad   \quad &
f_D \equiv \begin{pmatrix} \xi_D \\ \xi_{\bar{D}}^\dag \end{pmatrix}, 
    \quad  \text{and} \quad  f_a^c = i \gamma_0 \gamma_2 f_a^T \\  
  \mathcal{L}_m &=  \sum_i \: m_i {\bar f_i}  f_i \; , 
\end{split}
\label{dirac-spinor} 
\end{equation}
where $i$ runs over the flavor indices $\{ U,D \}$.  
In this basis, the four component Dirac spinors constructed 
are in the Weyl basis, which makes chirality manifest and is the 
most convenient choice for representing a relativistic chiral theory. 
In a non-relativistic theory, however, the Dirac basis is more suitable. 
The $\gamma$ matrices and all the four-component spinors can be rotated 
from the Weyl basis to the Dirac basis by the transformation
\begin{equation}
\gamma_\mu \rightarrow U  \gamma_\mu U^\dag  \quad \text{and} \quad
f_i \rightarrow U f_i \; ,  \quad \text{where} \quad 
U = \frac{1}{\sqrt{2}} \begin{pmatrix} 
					1 &  1 \\
					-1 &  1 
					\end{pmatrix} \; .		   
\label{dirac-spinor-dirac}
\end{equation}
The advantage of this basis is that $\gamma_0 = {\rm diag}(1,1,1,1)$ 
is diagonal, and so the dominant component of the Dirac four spinor 
(namely, $\left( 1 + \gamma_0 \right) f_i$), is a two component 
Pauli spinor, which furnishes the minimal representation for 
the non-relativistic spinor field. 
 
The next step is to eliminate the quirk mass scales $m_i$ while 
keeping the heavy quirk fields. This is accomplished by a simple 
time-dependent rescaling of the Dirac fermions in 
Eq.~\eqref{dirac-spinor-dirac}.  There is now a preferred frame, 
namely the center of mass frame, which is the frame in which we work
from now on.  Hence, the dominant component of the full Dirac spinors 
become
\begin{eqnarray}
  \psi_i &=& e^{ i m_i t} \  \frac{1}{2} \left( 1 + \gamma_0 \right) 
  f_i \label{nrspinor1}  \\
  \chi_i &=&  e^{- i m_i t} \  \frac{1}{2} \left( 1 + \gamma_0 \right) 
  f_i^c \; . \label{nrspinor2}
\end{eqnarray}
The field $\psi$ annihilates a heavy quirk field, while $\chi$
creates a heavy anti-quirk field.  These spinors roughly
correspond to particle and antiparticle and are appropriate for
a non-relativistic approximation about the center-of-mass frame 
of reference.  Classically, the entire quirk and the anti-quirk 
Dirac spinors in Eq.~\eqref{dirac-spinor} can be written in terms 
of the spinors in Eqs.~\eqref{nrspinor1} and \eqref{nrspinor2}: 
\begin{eqnarray}
f_i = e^{- i m_i t} 
	\begin{pmatrix} 
		\psi_i \\  \frac{ i \vec{\sigma} \cdot \vec{D}}{2 m_i + i D_0} \,  \psi_i 
	\end{pmatrix}  
\quad \text{and} \quad 	
f^c_i = e^{ i m_i t} 
	\begin{pmatrix} 
		\chi_i \\  \frac{ i \vec{\sigma} \cdot \vec{D}}{2 m_i + i D_0}\,  \chi_i 
	\end{pmatrix}   
\end{eqnarray}

The non-relativistic Lagrangian is written in terms of the spinors
$\psi$ and $\chi$ that designate almost on-shell quirks and
anti-quirks.  The Lagrangian can now be computed as an expansion in
$1/m_i$~\cite{Caswell:1985ui}:  
\begin{eqnarray}
 \mathcal{L}_0^\text{NR} &=& \sum_i \psi^\dag_i \left( 
    i D_0 + \frac{1}{2 m_i} \vec{D}^2 \right) \psi_i  
     + \frac{c_F g_q}{2 m_i}\: \psi_i^\dag \vec{\sigma}\cdot \vec{B}_q  
     \psi_i  \nonumber \\
       & & + \frac{  g_q c_D}{8 m_i^2} \psi^\dag_i  \left( 
     \vec{D} \cdot \vec{E}_q - \vec{E}_q \cdot \vec{D}\right) \psi_i +  
    \frac{  g_q c_s}{8 m_i^2} \psi^\dag_i \vec{\sigma}\cdot \left( 
     \vec{D} \times \vec{E}_q - \vec{E}_q \times \vec{D} \right)\psi_i 
       \label{nr-lagran} \\ \nonumber
       & & - \: \left( \psi_i \leftrightarrow \chi_i \right) + 
            \mathcal{O}\left( \frac{1}{m^3} \right)\; ,
\end{eqnarray}
where $i D_0 = i \partial_0 - g_q A_{q_0}$,   $i \vec{D} = i \vec{\nabla} +
i g_q \vec{A}_q$. 
The electric and magnetic quirky gauge fields are
defined as usual  $E_{q_i} = F_{q_{0i}}$ and 
$B_{q_i} = \epsilon_{ijk} F_q^{jk}$. 
At tree level, the matching is simply $c_F = c_D = c_s = 1$.
This is corrected due to quantum effects by
$\mathcal{O}(g_q^2)$.  The terms in this Lagrangian in 
Eq.~\eqref{nr-lagran} have well known physical interpretation. The 
$\vec{\sigma}\cdot \vec{B}_q $ term is the quirkcolor-magnetic moment 
interaction, the $\vec{D} \cdot \vec{E}_q$ term is the Darwin term and the 
$\vec{D} \times \vec{E}_q$ term is the spin-orbit coupling. The $1/m^3$ 
term contains the first relativistic correction.

\subsection{Spectrum}

The ground state spectrum of baryons and mesons can be determined by 
cataloging the Lorentz invariant bilinears made out of two $\psi$'s 
or one $\psi$ and one $\chi$, respectively.  The decomposition of the 
fermion bilinears in (flavor,spin) space can be written as
\begin{eqnarray}
(\mathbf{2},\mathbf{2}) \otimes (\mathbf{2},\mathbf{2})
& = &
(\mathbf{1}_a,\mathbf{1}_a) \oplus 
(\mathbf{3}_s,\mathbf{1}_a) \oplus
(\mathbf{1}_a,\mathbf{3}_s) \oplus
(\mathbf{3}_s,\mathbf{3}_s) \; ,
\label{represenation-eq}
\end{eqnarray}
where the subscripts $s$ and $a$ denote the symmetric and
antisymmetric representation, respectively.
Mesons can be written in all of these representations since they
are formed of non-identical particles.  
Baryons, however, must satisfy the Pauli exclusion principle.
Combining two identical quirks into a Lorentz invariant requires
an antisymmetric $SU(2)_Q$ contraction, and consequently,
only totally symmetric combinations of (flavor,spin) are possible,
specifically, $(\mathbf{1}_a,\mathbf{1}_a)$ and 
$(\mathbf{3}_s,\mathbf{3}_s)$.  The resulting baryonic states
can be written as 
\begin{eqnarray}
(\mathbf{1}_a,\mathbf{1}_a): & \qquad B^0_0  & = \;
   \tilde{\psi}^{i\alpha} \psi_{i\alpha} \\
(\mathbf{3}_s,\mathbf{3}_s): & \quad B^{F S} & = \; 
              \tilde{\psi}^{i \alpha}  
              (\sigma^F)_{i}^{\ j} (\sigma^S)_{\alpha}^{\ \beta} 
              \psi_{j \beta} \; ,
\end{eqnarray}
in terms of 
$\tilde{\psi}^{i\alpha} \equiv \epsilon^{ij}\epsilon^{\alpha\beta} \psi^T_{j\beta}$.
Here, flavor indices are designated by Latin letters ($i, j$, etc.), 
spin by Greek indices ($\alpha,\beta$, etc.), and quirkcolor
indices have been suppressed.  
The resulting states consist of: 
$B^0_0$, a complex scalar with zero electric charge and $+1$ baryon charge,
and $B^{FS}$, a massive spin-1 vector that is triplet under flavor
and also carries $+1$ baryon charge.  The anti-baryons with opposite
baryon charge are similarly constructed.  The $B^{FS}$ can be 
decomposed in terms of spin-1 baryons with electric charge $q$, 
denoted by $B^q_1$ (suppressing the spin index $S$), as 
\begin{eqnarray}
B^0_1     &\equiv& B^{3S} \\
B^{\pm}_1 &\equiv& \frac{1}{\sqrt{2}} \left( B^{1S} \mp i B^{2S} \right)
\label{charge-state-eq}
\end{eqnarray}

The mesons are made of $\chi$ and $\psi$ in representations given by
Eq.~(\ref{represenation-eq}).  In rough analogy to mesons in QCD
with two flavors, we can write the mesons as
\begin{eqnarray}
(\mathbf{1}_a,\mathbf{1}_a): \qquad \eta &=& 
   \tilde{\psi} \chi \\
(\mathbf{3}_s,\mathbf{1}_a): \qquad \pi  &=& 
   \tilde{\psi} \sigma^{F} \chi \\
(\mathbf{1}_s,\mathbf{3}_a): \qquad \phi &=& 
   \tilde{\psi} \sigma^{S} \chi \\
(\mathbf{3}_s,\mathbf{3}_s): \qquad \rho &=& 
   \tilde{\psi} \sigma^{F} \sigma^{S} \chi \; .
\end{eqnarray}
We have suppressed all fermion indices.  The proper combinations
of flavor states that yield definite electric charge states
are formed analogously to Eq.~(\ref{charge-state-eq}).

\subsection{Non-relativistic bound states in Quantum Quirky Dynamics}

We assume that the quirkcolor force is weakly coupled and thus 
described by perturbative physics with the quirks much heavier
than the resulting bound state energies.
In this limit, the timescale of the relative heavy quirk movement becomes 
much larger than the timescale of the quirky gluon dynamics. 
Then, feedback effects of the moving heavy quirks on gluons 
can be neglected, and so the adiabatic approximation or the leading
Born-Oppenheimer approximation should be applicable.  Also, one can use 
the different energy scales just like in the positronium problem in QED
(discarding the annihilation effect).  The non-relativistic bound state 
is characterized by the scale of the quirk mass $m$ (hard), 
the scale of the momentum transfer $p \sim mv$ (soft) 
and the scale of kinetic energy of the quirks in the 
center of mass frame $E \sim p^2/m \sim mv^2$ (ultrasoft), 
where $v$ is the heavy quirk velocity in this frame. 
In our weakly coupled non-relativistic system, 
$v \sim \alpha_q \ll 1$ and it follows that the three relevant scales are 
hierarchically ordered ({\it i.e.} $m \gg mv \gg mv^2$).  

Interestingly, such a situation also happens in quarkonium physics, 
where the same hierarchy has been utilized to construct equivalent 
effective theories to describe the quarkonium spectra and interactions
(for some reviews, see e.g.\ 
\cite{Pineda:1997bj,Brambilla:1999qa,Brambilla:1999xf,Eides:2000xc,Recksiegel:2003fm,Kniehl:2003ap,Brambilla:2004jw,Karshenboim:2005iy,Eichten:2007qx}).
First, the transition is from the EFT with relativistic quarks to
the non-relativistic effective theory (NREFT) with the quarks 
(and larger momenta) integrated out.  The theory describes dynamics 
of heavy quirk-antiquirk pairs at energy scales in the center-of-mass 
frame much smaller than their masses.  In Eq.~\eqref{nr-lagran}, 
we have reproduced the NREFT Lagrangian.  In quarkonium physics,  
a higher degree of simplification has been achieved by exploiting 
$mv \gg mv^2$ and building the so-called potential-NREFT (or pNREFT)
\cite{Pineda:1997bj,Brambilla:1999xf}, where degrees of freedom of 
$\sim mv$ are integrated out.  In this way, an analytical calculation 
of the spectrum becomes possible.  The ultra-soft degrees of freedom 
that remain dynamical in this theory are quirks of momentum $mv$
and energy $mv^2$ and quirky gauge fields of momentum and energy less
than $mv^2$.  The matching of the pNREFT to the NREFT 
is perturbative as long as $mv^2 \gtrsim \Lambda_{Q}$. 

The low energy theory is described in terms of quirk bilinears, which
depend on the relative distance between the two quirks,
$\vec{r} \equiv \vec{x}_1 - \vec{x}_2$, and the center-of-mass 
coordinate $\vec{R} \equiv (\vec{x}_1 + \vec{x}_2)/2$. 
All gauge fields are multipole expanded in $\vec{r}$, and therefore
depend only on $R$\@.  At leading order in the multipole 
expansion~\cite{Brambilla:1999qa,Brambilla:1999xf}, 
\begin{eqnarray}
  \mathcal{L} &=& S^\dag \left( i \partial_t + \frac{1}{2
      \mu} \vec{\partial}_r^{2} - V_s(r) 
           \right) S  \; .
\end{eqnarray}
In the above, $S$ is any quirkcolor singlet field and $V_s(r)$ is the
matching potential.  The reduced mass of the system is denoted by
$\mu$.  For our model, 
\begin{equation}
  \label{eq:red-mass}
  \frac{1}{\mu} = \frac{1}{m_U} + \frac{1}{m_D} \; .
\end{equation}
Hence, at leading order in the multipole expansion, the equation 
of motion of the singlet field is simply the Schr\"odinger equation!
Determining the bound state energies is thus very similar to a standard 
quantum mechanical calculation.  The main difference is that the
potential depends on a scale-dependent quirkcolor coupling that 
introduces $\log r$ dependence in the potential.

\refstepcounter{section}
\section*{Appendix~\thesection:~~UD-Parity}
\label{UD-parity-app}

In this appendix we demonstrate the charge radius operator is odd under 
a certain $Z_2$ symmetry under which the $U$-quirk transforms to $D$-quirk,
which we call ``UD-parity''.  Imposing that the ultraviolet theory
is UD-parity symmetric therefore eliminates the charge radius operator
and automatically ensures the lightest baryon is electrically neutral.

Consider the limit $m_U = m_D$.  In the ultraviolet theory,
one can show that the Lagrangian is symmetric under the 
following transformation: 
\begin{equation}
  \label{Z2-uv}
  \begin{array}{ccc}
  \xi_U \rightarrow \xi_D  \;\;  \text{and} \;\; 
  \xi_{\bar U} \rightarrow \xi_{\bar D}  
          & \quad \Rightarrow \quad &
  \psi_U \rightarrow \psi_D  \;\; \text{and} \;\; \chi_U \rightarrow \chi_D \\
  A^\mu_q \rightarrow A^\mu_q & \quad \text{and} \quad &
  A^\mu \rightarrow - A^\mu  
  \end{array}
\end{equation}
In the above $A_q^\mu$ and $A^\mu$ are the quirkcolor gauge fields and the 
electromagnetic gauge fields respectively.  
Under UD-parity, the electrically neutral scalar baryon is odd 
and the vector baryon is even, 
\begin{equation}
B_0    \rightarrow - B_0    \qquad 
B^{AI} \rightarrow   B^{AI} 
\end{equation}
and thus
\begin{equation}
E_i   \rightarrow - E_i   \qquad 
B_i   \rightarrow - B_i   \qquad 
r_D^2 \rightarrow - r_D^2 \; .
\end{equation}
Hence, $r_D^2$ must vanish if the low energy theory preserves
UD-parity.

\refstepcounter{section}
\section*{Appendix~\thesection:~~Integration Results}
\label{PS-app}

We have verified the results in \cite{Profumo:2006im} including
their Appendix, except for the expression for their $c_g$ in (A.4).  
We find the correct expression is
\begin{eqnarray}
c_g &=& \frac{2 \Delta M^2}{M_* \Gamma_{*}} 
+ \frac{2 M^2}{M_* \Gamma_{*}} 
  \frac{\Delta M^2 - m_*^2 \Gamma_{*}^2}{\Delta M^2 + m_*^2 \Gamma_{*}^2}
\end{eqnarray}
where $\Delta M^2 \equiv M_*^2 - M^2$.
With this correction, we were able to reproduce Ref.~\cite{Profumo:2006im}'s
numerical results (including their Fig.~3).

\end{appendix}

\section*{Acknowledgments}

We thank R.~Essig, R.~Harnik, M.~Luty, A.~Nelson, L.~Strigari, 
and J.~Wacker for helpful discussions at various stages of this work.
We also thank 
the Aspen Center for Physics, 
the Yukawa Institute of Theoretical Physics, 
the Kavli Institute of Theoretical Physics, and
the Galileo Galilei Institute for Theoretical Physics 
for hospitality where part of this work was completed.
This work was supported in part by the US Department of Energy 
under contracts DE-FG02-96ER40969 (GDK, TSR), 
DE-FG03-91ER40674 and DE-FG02-95ER40896 (KMR), DE-FG02-91ER40674 (JT), 
and by the NSF under contract PHY-0918108 (GDK).


\end{document}